\documentclass[a4paper,onecolumn,11pt,unpublished,noarxiv,allowtoday]{quantumarticle}
\pdfoutput=1
\usepackage[utf8]{inputenc}
\usepackage[english]{babel}
\usepackage[T1]{fontenc}
\usepackage{amsmath}
\usepackage{hyperref}
\usepackage{mathtools, nccmath}
\usepackage{setspace}
\usepackage{hyperref}
\usepackage{outlines}
\usepackage{fancyvrb}
\usepackage{mathtools}
\usepackage{physics}
\usepackage{comment}
\usepackage{bbold}
\usepackage{amssymb}
\usepackage{cancel}
\usepackage{graphicx}
\usepackage[caption=false]{subfig}
\usepackage[usenames, dvipsnames]{color}
\usepackage{soul}
\usepackage{tensor}
\usepackage{amsthm}
\usepackage[caption=false]{subfig}
\usepackage{algorithm}
\usepackage{algcompatible}

\usepackage[numbers,sort&compress]{natbib} 
\usepackage{tikz}
\usepackage{lipsum}
\usepackage{tkz-graph}
\usetikzlibrary{hobby,backgrounds,calc,trees}
\usepackage{xcolor}
\usepackage{soul}
\newcommand{\mathcolorbox}[2]{\colorbox{#1}{$\displaystyle #2$}}
\usetikzlibrary{positioning}
\usepackage{yquant}
\usepackage{caption}
\usepackage{subfig}
\useyquantlanguage{groups}

\usepackage{makecell}

\usepackage{tikz,esvect,tikz-3dplot}
\usetikzlibrary{3d,calc,intersections}
\usepackage{tikz}
\usepackage{tikz-3dplot}

\usepackage{adjustbox}

\pgfmathsetmacro\MathAxis{height("$\vcenter{}$")}

\usepackage{datetime}

\begin{document}

\title{Simple magic state calculations using `Improved Simulation of Stabilizer Circuits'}
\date{\today}

\author{Kwok Ho Wan}
\orcid{0000-0002-1762-1001}
\affiliation{Universal Quantum, Gemini House, Mill Green Business Estate, Haywards Heath, UK, RH16 1XQ}
\thanks{current affiliation: Academic Visitor, Blackett Laboratory, Imperial College London, South Kensington, London SW7 2AZ, UK}
\email{kwok((dot))wan14((at))imperial.ac.uk}

\begin{abstract}
We employed the techniques from [Phys. Rev. A \textbf{70}, 052328 (2004)/arXiv:0406196] to analytically study two set of quantum circuits containing one $T$ gate/magic $\ket{T} = \frac{\ket{0}+\sqrt{i}\ket{1}}{\sqrt{2}}$ state. These include the $T$ state via gate teleportation and magic state injection on the rotated surface code.
\end{abstract}

\maketitle
\section{Introduction}
Magic $\ket{T} = \frac{\ket{0}+\sqrt{i}\ket{1}}{\sqrt{2}}$ states are vital resource states in fault-tolerant quantum computation \cite{gidney2024magicstatecultivationgrowing}. In this working paper, we follow the techniques from \cite{Aaronson_2004} to accommodate for extended stabiliser simulations involving one magic state\footnote{We shall interchange the Aaronson-Gottesman approach \cite{Aaronson_2004} and the term \textit{extended stabiliser simulations} synonymously throughout this text.}. We will use this technique to explicitly study:
\begin{enumerate}
    \item the consumption of a magic state to perform a $T = \begin{pmatrix}
       1 & 0 \\
       0 & \sqrt{i}
    \end{pmatrix}$ gate, and 
    \item magic state injection on the rotated surface code \cite{LaoInjection2022}.
\end{enumerate}
We start with a quick review of magic states, followed by Aarsonson-Gottesman's improved (extended) stabiliser simulation technique \cite{Aaronson_2004}, before moving onto the simple magic state calculations.

\section{Consumption of $\ket{T}$ state}
The magic $\ket{T}$ state is a fixed angle ($\theta = \pi/4$) special case of a general class of states, $\ket{m_{\theta}}$:
\begin{equation}
    \ket{m_{\theta}} = \frac{1}{\sqrt{2}} \Big( \ket{0} + e^{i\theta}\ket{1} \Big) \ .
\end{equation}
A single $\ket{m_{\theta}}$ state can be consumed (measurement binary result $\alpha \in \{0,1\}$) to perform a rotational gate along the $Z$ axis by gate teleportation.
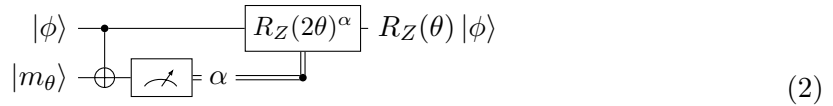
\begin{equation}
    \begin{tikzpicture}
    \begin{yquant}

    qubit {$\ket{\phi}$} q[+1];

    qubit {$\ket{m_{\theta}}$} q[+1];
    
    cnot q[1] | q[0];
        
    measure q[1]; text {$\alpha$} q[1];
    box {\small$R_Z(2\theta)^\alpha$} (q[0]) | q[1];
    discard q[1];
    output {$R_Z(\theta)\ket{\phi}$} q[0];
    \end{yquant}
    \end{tikzpicture}
    \label{eq:rz_gate_by_teleportation}
\end{equation}
The rotational $Z$ gate, $R_{Z}(\theta)$ in the computational basis is:
\begin{equation}
    R_{Z}(\theta) = 
    \begin{pmatrix}
       1 & 0 \\
       0 & e^{i\theta}
    \end{pmatrix} \ .
\end{equation}
For an arbitrary angle $\theta$, the byproduct correction is another non-Clifford operator $R_Z(2\theta)$ when the measurement result $\alpha=1$. In the context of consuming magic states, we fixed the angle $\theta$ to be $\pi/4$, the $T =     \begin{pmatrix}
       1 & 0 \\
       0 & \sqrt{i}
    \end{pmatrix}$ gate can be performed with a $S=\begin{pmatrix}
       1 & 0 \\
       0 & i
    \end{pmatrix}$ Clifford gate byproduct correction: 
\begin{equation}
\centering
    \begin{tikzpicture}
    \begin{yquant}
    qubit {$\ket{{\psi}}$} q[+1];
    qubit {$\ket{T}$} a[+1];
    
    CNOT a[0] | q[0];
  
    measure a[0];
    text {$\alpha$} a[0];
    box {$S^\alpha$} (q[0]) | a[0];
    discard a[0];
    output {$T\ket{\psi}$} q[0];
  
    \end{yquant}
    \end{tikzpicture} \ .
    \label{eq:magic_consume}
\end{equation}
The ability to consume (offline produced) magic states to perform a $T$ gates by teleportation forms the foundation of surface code fault-tolerant quantum computation \cite{fowler2019lowoverheadquantumcomputation} as there are stringent restrictions on performing transversal gates on quantum codes \cite{eastin_knill_2009}. We shall describe the limitations to simulating circuits with $T$ gates in the next section.

\section{Simulation of non-Clifford quantum circuits}
The consumption of a magic $\ket{T}$ state to perform a $T$ gate is a non-Clifford \cite{gottesman2009introductionquantumerrorcorrection} quantum circuit (equation \ref{eq:magic_consume}). We cannot rely on Clifford frame simulators such as Stim \cite{gidney2021stim}. For a $n$ qubit quantum circuit with $t$ number $T$ gates, the classical time complexity of simulation is $\mathcal{O}(2^{t}\text{poly}(n))$ \cite{gottesman1998heisenbergrepresentationquantumcomputers,Bravyi_Gosset_2016_PRL,nakhl2024stabilizertensornetworksmagic}. It is also highly unlikely for efficient classical simulations of circuits with $t > \mathcal{O}(n)$ $T$ gates to exist, given strong complexity-theoretical conjectures \cite{Bremner_2010}. Although better classical hardware \cite{Menczer_NVIDIA_2024} and simulation methods \cite{nakhl2024stabilizertensornetworksmagic,bayraktar2023cuquantumsdkhighperformancelibrary} are consistently pushing the simulation boundaries on the size of non-Clifford circuits. For example, tensor network simulations for $n=4000$ qubits and $t = 320$ $T$ doped circuits are possible \cite{nakhl2024stabilizertensornetworksmagic}. These numbers can potentially be pushed further up with dedicated tensor contraction hardware, such as the NVIDIA GH200 Grace Hopper Superchip \cite{Menczer_NVIDIA_2024,bayraktar2023cuquantumsdkhighperformancelibrary}.

In general, a $n$-qubit non-Clifford state can be decomposed as a linear combination of $\chi$ number of stabiliser states \cite{Aaronson_2004,Bravyi_Gosset_2016_PRL,Bravyi_2019}:
\begin{equation}
    \ket{\Psi} = \sum_{k=1}^{\chi} \lambda_k \ket{\psi_k} = U \ket{0}^{\otimes n} \ ,
\end{equation}
where $\ket{\psi_k}$ are Clifford states and $\chi$ is commonly known as the stabiliser rank. In general, $\chi = \mathcal{O}(2^{t})$, where $t$ is the number of $T$ gates involved in constructing the unitary $U$. By fixing $t=1$, we hope to decorate stabiliser frames simulators (such as Stim) with stabiliser decomposition techniques \cite{Aaronson_2004,Bravyi_Gosset_2016_PRL,Bravyi_2019} to simulate a low magic non-stabiliser state. 

\section{Aaronson-Gottesman's approach}
We shall now review Aaronson-Gottesman's approach \cite{Aaronson_2004} to perform extended stabiliser calculations involving non-Clifford gates with Clifford input states. This is practically relevant to magic state injection \cite{LaoInjection2022,Li_2015}, as only one physical $T$ gate is performed in the whole injection protocol. Hence, a finite number of stabiliser terms need to be tracked in order to perform this non-Clifford calculation. We shall illustrate how to manipulate general non-Clifford stabiliser decomposition simulation through the Aaronson-Gottesman approach below, before focusing on a single $T$ gate applied at the start of a Clifford circuit afterwards.
\subsection{General stabiliser decomposition (mixed state) \cite{Aaronson_2004}}
Suppose we have the stabiliser state: 
\begin{equation}
    \rho = \frac{1}{2^r}\prod_{j=1}^{r}(I+g_j) \ ,
\end{equation}
stabilised by stabiliser generators:
\begin{equation}
    \langle g_1,g_2,...,g_r\rangle \ .
\end{equation}
This arbitrary Clifford state can be generated by the circuit below (equation \ref{eq:clifford_cir_general}):
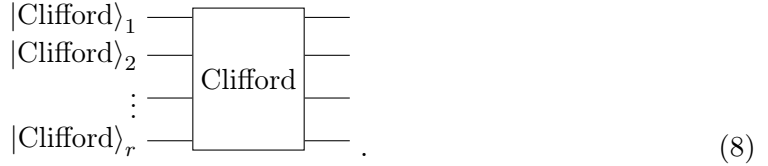
\begin{equation}
\label{eq:clifford_cir_general}
\centering
    \begin{tikzpicture}
    \begin{yquant}
    qubit {$\ket{\text{Clifford}}_1$} q[+1];
    qubit {$\ket{\text{Clifford}}_2$} q[+1];
    qubit {$\vdots$} q[+1];
    qubit {$\ket{\text{Clifford}}_{r}$} q[+1];
    hspace {0.5cm} - ;
    box {$\text{Clifford}$} (q[0]-q[3]);
    hspace {0.5cm} - ;
    \end{yquant}
    \end{tikzpicture} \ .
\end{equation}
Any arbitrary non-Clifford single or multi-qubit unitary operation, $U$, can be decomposed into a linear combination of single-/multi-qubit Pauli operators:
\begin{equation}
    U = \sum_{i}^{\nu} c_i P_i \ .
\end{equation}
The application of this unitary operator to the density matrix is given by: 
\begin{equation}
    U\rho U^{\dagger} = \frac{1}{2^r} \sum_{i}^{\nu} c_i P_i \prod_{j=1}^{r}(I+g_j) \sum_{k}^{\nu} c_k^{*} P_k \ .
\end{equation}
The state, $U\rho U^{\dagger}$ is constructed with the circuit (equation \ref{eq:clifford_cir_general_one_non_clifford}):
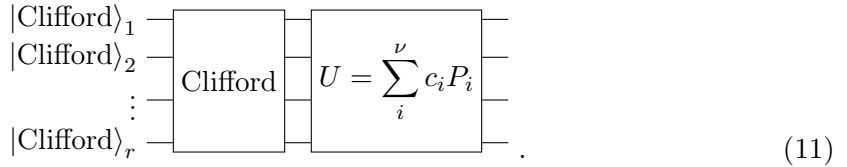
\begin{equation}
\label{eq:clifford_cir_general_one_non_clifford}
\centering
    \begin{tikzpicture}
    \begin{yquant}
    qubit {$\ket{\text{Clifford}}_1$} q[+1];
    qubit {$\ket{\text{Clifford}}_2$} q[+1];
    qubit {$\vdots$} q[+1];
    qubit {$\ket{\text{Clifford}}_{r}$} q[+1];
    hspace {0.25cm} - ;
    box {$\text{Clifford}$} (q[0]-q[3]);
    hspace {0.25cm} - ;
    box {$U=\displaystyle\sum_{i}^{\nu} c_i P_i $} (q[0]-q[3]);
    hspace {0.25cm} - ;
    \end{yquant}
    \end{tikzpicture} \ .
\end{equation}
We can commute (or anti-commute) the factor term $\sum_{k} c_k^{*} P_k$ through to the left of $\prod_{j=1}^{r}(1+g_j)$:
\begin{equation}
    U\rho U^{\dagger} = \frac{1}{2^r} \sum_{i}^{\nu}\sum_{k}^{\nu} c_ic_k^{*} P_iP_k \prod_{j=1}^{r}(I+(-1)^{\omega(g_j, P_k)}g_j) \ ,
\end{equation}
where $\omega(g_j, P_k)$ is the symplectic inner product between $g_j$ and $P_k$:
\begin{equation}
    \omega(A, B) = \begin{cases}
      0, & \text{if}\ [A, B]=0 \\
      1, & \text{if}\ \{A, B\}\neq 0
    \end{cases} \ ,
\end{equation}
the symplectic inner product, $\omega(A, B)$, returns $0$ when $A$ and $B$ commutes and returns $1$ when they anti-commute. Now, we can see that the density matrix is just a superposition of $\nu^2$ terms, where each term is itself a stabiliser state with a left multiplication of $P_{i}P_{k}$ to it. By linearity, we can apply the somewhat modified stabiliser state manipulation rules on top, decorated with the superposition of stabiliser states. Like Aaronson-Gottesman, we shall rename $c_ic_k^{*} = c_{i,k}$ $P_iP_k = P_{i,k}$,
\begin{equation}
    U\rho U^{\dagger} = \frac{1}{2^r} \sum_{i}^{\nu}\sum_{k}^{\nu} c_{i,k} P_{i,k} \underbrace{\prod_{j=1}^{r}(I+(-1)^{\omega(g_j, P_k)}g_j)}_{\text{Clifford state}} \ .
\end{equation}
Next, we shall outline the application of Clifford unitary or Clifford measurement to this state and observed the modified stabiliser transformation rules.

\subsection{Unitary operation}
Let's name the superposition of stabiliser states $\rho'$:
\begin{equation}
    \rho' = \frac{1}{2^r} \sum_{i}^{\nu}\sum_{k}^{\nu} c_{i,k} P_{i,k} \prod_{j=1}^{r}(I+(-1)^{\omega(g_j, P_k)}g_j) \ .
\end{equation}
Apply a Clifford unitary $U_{c}$ on $\rho'$, then 
\begin{equation}
\begin{split}
        U_c \rho' U_c^{\dagger} = \frac{1}{2^r} \sum_{i,k} & c_{i,k} U_c P_{i,k} U_c^{\dagger} \cdot U_c \prod_{j=1}^{r}(I+(-1)^{\omega(g_j, P_k)}g_j) U_c^{\dagger} \ .
\end{split}
\end{equation}
We will have to transfrom the $\nu^2$ Pauli operators via: 
\begin{equation}
    P_{i,k} \rightarrow U_c P_{i,k} U_c^{\dagger} \ ,
\end{equation}
and transform their corresponding stabiliser states (only $\nu$ of them) via:
\begin{equation}
    \prod_{j=1}^{r}(I+(-1)^{\omega(g_j, P_k)}g_j) \rightarrow U_c \prod_{j=1}^{r}(I+(-1)^{\omega(g_j, P_k)}g_j) U_c^{\dagger} \ ,
\end{equation}
which means you can just apply the standard stabiliser generator transformation rules for the $\nu$ different set of generators:
\begin{equation}
    U_c \ \Bigg\{\Bigg\langle \bigcup_{j=1}^{r} (-1)^{\omega(g_j, P_k)}g_j \Bigg\rangle\Bigg\}_{k=1}^{\nu} \ U_c^{\dagger} \ , 
\end{equation}
and also track the additional terms to the left of this in the stabiliser decomposed density matrix: 
\begin{equation}
    c_{i,k}P_{i,k} \rightarrow c_{i,k} U_c P_{i,k} U_c^{\dagger} \ ,
\end{equation}
\subsection{Pauli measurements}
If one measures in the Pauli operator $Q$ basis, then its projections are $\frac{1+Q}{2}$ or $\frac{1-Q}{2}$ depending on the parity measurement value. Name: $Q^{\pm} = 1 \pm Q$, the un-normalised projector (similar to how they are labelled in \cite{Aaronson_2004}). If we were to measure $\rho'$, $\rho' \rightarrow \frac{Q^{\pm}}{2} \rho' \frac{Q^{\pm}}{2}$, the resulting un-normalised density matrix is:
\begin{equation}\begin{split}
    \frac{Q^{\pm}}{2} & \rho'\frac{Q^{\pm}}{2} = \frac{1}{2^{r+2}} Q^{\pm}\sum_{i,k} c_{i,k} P_{i,k} \prod_{j=1}^{r}(I+(-1)^{\omega(g_j, P_k)}g_j)Q^{\pm} \ .
\end{split}\end{equation}
Let's go through the different cases of whether if $Q$ commutes with the stabiliser generators or anti-commutes with some stabiliser generators.

\subsubsection{$Q$ commutes with all $g_j$}
This means $Q$ is in the set of $\{g_j\}$.
\begin{equation}\begin{split}
    \frac{Q^{\pm}}{2} & \rho'\frac{Q^{\pm}}{2} = \rho'(\pm) = \frac{1}{2^{r+2}} \sum_{i,k} c_{i,k} Q^{\pm}P_{i,k}Q^{\pm} \prod_{j=1}^{r}(I+(-1)^{\omega(g_j, P_k)}g_j) \ .
\end{split}\end{equation}
If $Q$ commutes with $P_{i,k}$ then $Q^{\pm}P_{i,k}Q^{\pm} = 2P_{i,k}Q^{\pm}$, if $Q$ anti-commutes with $P_{i,k}$ then $Q^{\pm}P_{i,k}Q^{\pm} = 0$. Then, in general, 
\begin{equation}
    \rho'(\pm) = \frac{1}{2^{r+2}} \sum_{i\in A}\sum_{k \in B_{\pm}} c_{i,k} P_{i,k} \prod_{j=1}^{r}(I+(-1)^{\omega(g_j, P_k)}g_j) \ .
\end{equation}
where index $i$ is summed over set $A$ when $P_{i,k} = P_iP_k$ commutes with $Q$. Suppose $g_b \propto Q$, then the index $k$ is summed over set $B_{\pm}$:
\begin{equation}
    B_{\pm} = \{k \in B_{\pm} | (-1)^{\omega(g_b, P_k)}g_b = \pm Q \} \ .
\end{equation}
Note that the density matrix is not normalised.

\subsubsection{$Q$ anti-commutes with some $g_j$}
Suppose $Q$ anti-commutes with a set of generators: $\{g_1,g_2,g_3,...,g_n\}$, we can re-multiply stabiliser $g_1$ to every element in $\{g_2, ... ,g_n\}$, such that only $g_1$ anti-commutes with $Q$, the stabilisers modified by internal group multiplications are: $\{g_2' = g_1g_2,g_3' = g_1g_3,...,g_n' = g_1g_n \} $, which no longer anti-commute with $Q$. With this re-write of group generators, only $g_1$ anti-commutes with $Q$. The density matrix after the measurement is:
\begin{equation}\begin{split}
\rho'(\pm) = \frac{1}{2^{r+2}} \sum_{i,k} & c_{i,k} Q^{\pm}P_{i,k} \cdot \Big[(I+(-1)^{\omega(g_1, P_k)}g_1)Q^{\pm}\Big]\Lambda_k \ ,
\end{split}\end{equation}
where $\Lambda_k$ is given by:
\begin{equation}\begin{split}
\Lambda_k = \prod_{j=2}^{r}(I+(-1)^{\omega(g_j, P_k)}g_j) \ .
\end{split}\end{equation}
With some modifications:
\begin{equation}\begin{split}
\rho'(\pm) = \frac{1}{2^{r+2}} \sum_{i,k} & c_{i,k} Q^{\pm}P_{i,k} \cdot \Big[(Q^{\pm}+(-1)^{\omega(g_1, P_k)}Q^{\mp}g_1)\Big]\Lambda_k \ ,
\end{split}\end{equation}

\begin{enumerate}
    \item If $P_{i,k}$ commutes with $Q$,
    \begin{equation}
        \label{eq:update_rule_1}
        \rho'(\pm) = \frac{1}{2^{r+2}} \sum_{i,k} c_{i,k} P_{i,k}(2 Q^{\pm})\Lambda_k \ .
    \end{equation}
    \item If $P_{i,k}$ anti-commutes with $Q$,
    \begin{equation}
    \begin{split}
        \rho'(\pm) = \frac{1}{2^{r+2}} \sum_{i,k} &(-1)^{\omega(g_1, P_k)} (-1)^{1+\omega(Q, P_{i,k})} c_{i,k} P_{i,k} (I\delta_{\omega(Q, P_{i,k}),0}+g_1\delta_{\omega(Q, P_{i,k}),1})) (2 Q^{\pm})\Lambda_k \ ,
        \end{split}
    \end{equation} 
    where $\delta_{i,j}$ is the Kronecker delta. In other words, \textbf{for the specific anti-commuting $P_{i,k}$ terms}, 
    \begin{equation}
        \label{eq:update_rule_2}
        \begin{split}
            c_{i,k} & \rightarrow (-1)^{\omega(g_1, P_k)} c_{i,k} \ , \\
            P_{i,k} & \rightarrow P_{i,k} g_1 \ .
        \end{split}
    \end{equation}
\end{enumerate}
Any un-normalized density matrix can be normalized by dividing it by dividing it by normalisation factor $\text{tr}(\rho'(\pm))$. In summary, throughtout the entire extended stabiliser simulation, the objects to track are:
\begin{enumerate}
    \item  $c_{i,k}P_{i,k}$,
    \item $\displaystyle\Bigg\{\Bigg\langle \bigcup_{j=1}^{r} (-1)^{\omega(g_j, P_k)}g_j \Bigg\rangle\Bigg\}_{k=1}^{\nu}$ and
    \item the trace of the density matrix computed at the end.
\end{enumerate}
Let's try to apply these modified stabiliser decomposition manipulation rules to circuits with a single $T$ gate, a common theme in most non-Clifford state initialisation schemes \cite{Li_2015,LaoInjection2022,gidney2024magicstatecultivationgrowing}

\section{Non-Clifford circuit with a single $T$ gate}
For non-Clifford circuits with a single non-Clifford rotation gate at the beginning of the circuit, we need to decompose a $R_{Z}(\theta)$ gate into a linear combination of Pauli gates first. This is given by:
\begin{equation}
\begin{split}
        R_Z(\theta) & = \frac{1+e^{i\theta}}{2}I + \frac{1-e^{i\theta}}{2}Z \\
        & = e^{i\theta/2} \big(\text{cos}(\theta/2)I-i \ \text{sin}(\theta/2)Z\big)
\end{split}
\end{equation}
where $I$ is the qubit identity operator. Suppose we have an initial set of $r$ stabiliser generators $G$:
\begin{equation}
    G = \langle g_1, ... , g_{l-1}, g_l = X_l, g_{l+1}, ... , g_r\rangle \ ,
\end{equation}
this state is constructed with the following circuit in equation \ref{eq:clifford_cir}.
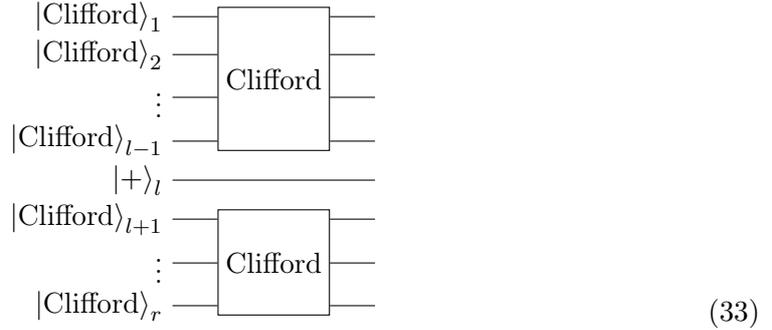
\begin{equation}
\label{eq:clifford_cir}
\centering
    \begin{tikzpicture}
    \begin{yquant}
    qubit {$\ket{\text{Clifford}}_1$} q[+1];
    qubit {$\ket{\text{Clifford}}_2$} q[+1];
    qubit {$\vdots$} q[+1];
    qubit {$\ket{\text{Clifford}}_{l-1}$} q[+1];
    qubit {$\ket{+}_{l}$} q[+1];
    qubit {$\ket{\text{Clifford}}_{l+1}$} q[+1];
    qubit {$\vdots$} q[+1];
    qubit {$\ket{\text{Clifford}}_r$} q[+1];
    hspace {0.5cm} - ;
    box {$\text{Clifford}$} (q[0]-q[3]);
    box {$\text{Clifford}$} (q[5]-q[7]);
    hspace {0.5cm} - ;
    \end{yquant}
    \end{tikzpicture}
\end{equation}
This is stabilised by state (density matrix $\rho$):
\begin{equation}
    \rho = \frac{1}{2^r}\prod_{j=1}^{l-1}(I+g_j) \ (I+X_l) \ \prod_{j=l+1}^{r}(I+g_j) \ .
\end{equation}
Suppose we apply a single $R_Z(\theta)$ gate according to this circuit in equation \ref{eq:clifford_cir_single_RZ}:
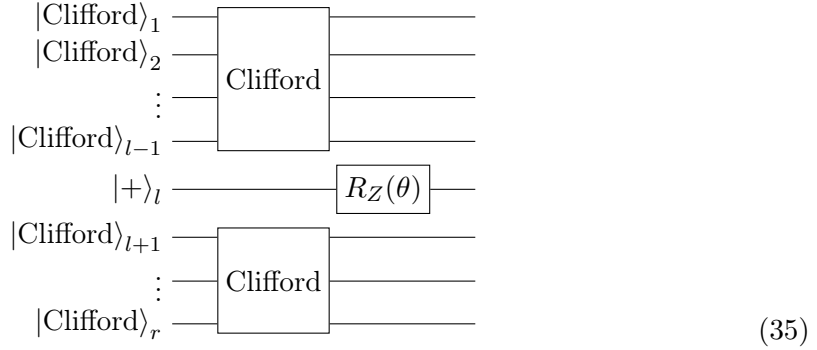
\begin{equation}
\label{eq:clifford_cir_single_RZ}
\centering
    \begin{tikzpicture}
    \begin{yquant}
    qubit {$\ket{\text{Clifford}}_1$} q[+1];
    qubit {$\ket{\text{Clifford}}_2$} q[+1];
    qubit {$\vdots$} q[+1];
    qubit {$\ket{\text{Clifford}}_{l-1}$} q[+1];
    qubit {$\ket{+}_{l}$} q[+1];
    qubit {$\ket{\text{Clifford}}_{l+1}$} q[+1];
    qubit {$\vdots$} q[+1];
    qubit {$\ket{\text{Clifford}}_r$} q[+1];
    hspace {0.5cm} - ;
    box {$\text{Clifford}$} (q[0]-q[3]);
    box {$\text{Clifford}$} (q[5]-q[7]);
    align  q;
    box {$R_Z(\theta)$} q[4];
    hspace {0.5cm} - ;
    \end{yquant}
    \end{tikzpicture}
\end{equation}
The output state is (density matrix $\rho'$):
\begin{equation}
    \begin{split}
        \rho' & = R_Z(\theta)_l \rho R_Z(\theta)^{\dagger}_l \\ 
        & =  \sum_{i}^{\nu}\sum_{k}^{\nu} c_{i,k} P_{i,k} (I+(-1)^{\omega(X_l, P_k)}X_l) \\ 
        & \ \ \ \ \ \cdot \frac{1}{2^r}\prod_{j\neq l}(I+g_j) \ ,
    \end{split}
\end{equation}
where $\nu = 2$ in this case, the following matrix, $D$, can be used to represent the $c_{i,k}P_{i,k}$ factor via $D_{i,k} = c_{i,k}P_{i,k}$:
\begin{equation}
    D = \begin{pmatrix}
       \text{cos}^2(\theta/2) I & (i/2)\text{sin}(\theta) Z_l \\ 
       -(i/2)\text{sin}(\theta) Z_l  & \text{sin}^2(\theta/2) I
    \end{pmatrix} \ ,
\end{equation}
the $i,k$ indices can range over $\{I,Z\}$ now.
Let's re-write $\rho'$:
\begin{equation}
    \begin{split}
        \rho' =  \sum_{i,k} D_{i,k} (I+(-1)^{\omega(X_l, P_k)}X_l) \frac{1}{2^r}\prod_{j\neq l}(I+g_j) \ ,
    \end{split}
\end{equation}
where $k=I: \ (-1)^{\omega(X_l, P_I)} = +1$ and $k=Z: \ (-1)^{\omega(X_l, P_Z)} = -1$.

\subsection{$\ket{T}$ state stabiliser decomposition representations}
For the $\ket{T}$ state, it can be generated via $R_Z(\pi/4)\ket{+}$ or $R_Z(-\pi/4)\ket{Y}$, where $\ket{Y} \propto \ket{0} + i \ket{1}$. Hence for $\theta = \pi/4$, $\rho'$ has two equivalent representations.

\begin{equation}
    \begin{split}
        \rho' & =  \sum_{i,k} D_{i,k} (I+(-1)^{\omega(X_l, P_k)}X_l) \frac{1}{2^r}\prod_{j\neq l}(I+g_j) \ , \\ 
        D & = \begin{pmatrix}
       \text{cos}^2(\pi/8) I & (i/2)\text{sin}(\pi/4) Z_l \\ 
       -(i/2)\text{sin}(\pi/4) Z_l  & \text{sin}^2(\pi/8) I
    \end{pmatrix} \ .
    \end{split}
\end{equation}

\begin{equation}
    \begin{split}
        \rho' & =  \sum_{i,k} D_{i,k} (I+(-1)^{\omega(Y_l, P_k)}Y_l) \frac{1}{2^r}\prod_{j\neq l}(I+g_j) \ , \\ 
        D & = \begin{pmatrix}
       \text{cos}^2(\pi/8) I & -(i/2)\text{sin}(\pi/4) Z_l \\ 
       (i/2)\text{sin}(\pi/4) Z_l  & \text{sin}^2(\pi/8) I
    \end{pmatrix} \ .
    \end{split}
\end{equation}
These density matrices represent the non-Clifford joint state $\ket{T}_{l}\otimes \ket{\text{Clifford}}_{1,2,3,...,j,...,n; j\neq l}$. 

\section{$T$ gate by teleportation}
Let's look at the $T$ gate by teleportation applied to the Clifford state $\ket{\psi} = \ket{+}$. Although this is a simple calculation that can be performed via the brute force state vector picture, it will be useful to perform this calculation with Aaronson-Gottesman's technique as a starter problem.
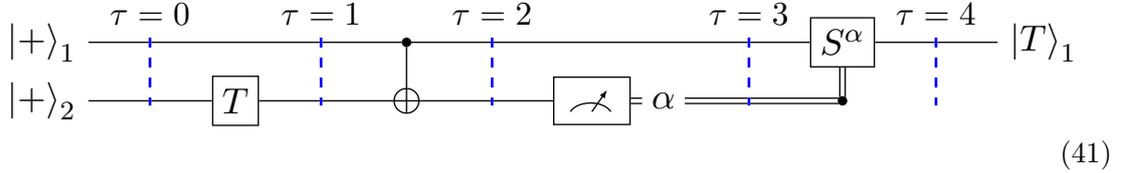
\begin{equation}
\centering
\label{eq:T_teleport_stab_try}
\resizebox{.95\linewidth}{!}{%
    \begin{tikzpicture}
    \begin{yquant}
    qubit {$\ket{{+}}_1$} q[+1];
    qubit {$\ket{+}_2$} a[+1];
    [blue, thick, label={\small$\tau = 0$}]
    barrier (-);
    box {$T$} a[0];
    [blue, thick, label={\small$\tau = 1$}]
    barrier (-);
    CNOT a[0] | q[0];
    [blue, thick, label={\small$\tau = 2$}]
    barrier (-);
    measure a[0];
    text {$\alpha$} a[0];
    [blue, thick, label={\small$\tau = 3$}]
    barrier (-);
    box {$S^\alpha$} (q[0]) | a[0];
    discard a[0];
    [blue, thick, label={\small$\tau = 4$}]
    barrier (-);
    output {$\ket{T}_1$} q[0];
    \end{yquant}
    \end{tikzpicture}%
    }
\end{equation}
At time $\tau = 1$, the state is:
\begin{equation}
    \begin{split}
        \rho' =  \sum_{i,k} D_{i,k} (I+(-1)^{\delta_{k,Z}}X_2) \frac{1}{2^2}(I+X_1) \ ,
    \end{split}
\end{equation}
\begin{equation}
    D = \begin{pmatrix}
       \text{cos}^2(\pi/8) I & (i/2)\text{sin}(\pi/4) Z_2 \\ 
       -(i/2)\text{sin}(\pi/4) Z_2  & \text{sin}^2(\pi/8) I
    \end{pmatrix} \ ,
\end{equation}
where $\delta_{k,Z}$ is the Kronecker delta tensor.
At time $\tau = 2$, the application of a CNOT gate leads to:
\begin{equation}
    \begin{split}
        \rho' =  \sum_{i,k} D_{i,k} (I+(-1)^{\delta_{k,Z}}X_2) \frac{1}{2^2}(I+X_1X_2) \ ,
    \end{split}
\end{equation}
\begin{equation}
    D = \begin{pmatrix}
       \text{cos}^2(\pi/8) I & (i/2)\text{sin}(\pi/4) Z_1Z_2 \\ 
       -(i/2)\text{sin}(\pi/4) Z_1Z_2  & \text{sin}^2(\pi/8) I
    \end{pmatrix} \ ,
\end{equation}
At time $\tau = 3$, the measurement in the $Q = Z_2$ basis commutes with all the $P_{i,k}$, but anti commutes with some $g_j$, namely $X_2$. This means $\Lambda_k = (1+(-1)^{\delta_{k,Z}}X_1)$.
\begin{equation}
    \begin{split}
        \rho' =  \sum_{i,k} D_{i,k} (I+(-1)^{\delta_{k,Z}}X_2) \frac{1}{2^2}(I+(-1)^{\delta_{k,Z}}X_1) \ ,
    \end{split}
\end{equation}
Post-measurement with parity measurement results: $(-1)^{\alpha}$,
\begin{equation}
    \begin{split}
        \Rightarrow & \rho'((-1)^\alpha) =  \frac{1}{2}\sum_{i,k} D_{i,k} (I+(-1)^{\delta_{k,Z}}X_1) (I +(-1)^\alpha  Z_2) \ ,
    \end{split}
\end{equation}
\begin{equation}
    D = \begin{pmatrix}
       \text{cos}^2(\pi/8) I & (i/2)\text{sin}(\pi/4) Z_1Z_2 \\ 
       -(i/2)\text{sin}(\pi/4) Z_1Z_2  & \text{sin}^2(\pi/8) I
    \end{pmatrix} \ .
\end{equation}
Let's look at $(I +(-1)^\alpha  Z_2)$, we can modify it to extract the $Z_2$ term and the phase $(-1)^\alpha$ as multiplicative factors: $(I +(-1)^\alpha  Z_2) = (-1)^\alpha Z_2(I+(-1)^\alpha Z_2)$. We can modify only the off-diagonal terms in $D$ with this trick and multiply the $(-1)^\alpha Z_2$ through to cancel the $Z_2$.
\begin{equation}
\begin{split}
        & \Rightarrow D = \begin{pmatrix}
       \text{cos}^2(\pi/8) I & (-1)^m(i/2)\text{sin}(\pi/4) Z_1 \\ 
       -(-1)^m(i/2)\text{sin}(\pi/4) Z_1  & \text{sin}^2(\pi/8) I
    \end{pmatrix} \ .
\end{split}
\end{equation}
Let's ignore the $ (I +(-1)^\alpha  Z_2)$ term in $\rho'((-1)^\alpha)$ for now, as it's just a separate state $\ket{\alpha \in \{ 0,1 \}}$ in the computational basis of subsystem $2$. Focusing on subsystem $1$ only (rename this as $\rho''$):
\begin{equation}
    \begin{split}
        & \rho''((-1)^\alpha) =  \frac{1}{2}\sum_{i,k} D_{i,k} (I+(-1)^{\delta_{k,Z}}X_1) \ , \\
        & D = \begin{pmatrix}
       \text{cos}^2(\pi/8) I & (-1)^m(\frac{i}{2})\text{sin}(\pi/4) Z_1 \\ 
       (-1)^{m+1}(\frac{i}{2})\text{sin}(\pi/4) Z_1  & \text{sin}^2(\pi/8) I
    \end{pmatrix} \ .
    \end{split}
\end{equation}
For binary measurement result: $\alpha=0$, this is exactly the $\ket{T}$ state. For $\alpha=1$, it's trickier. We need to perform a conditional $S$ gate as shown before $\tau = 4$ in equation \ref{eq:T_teleport_stab_try}. Note that $S: X\rightarrow Y, \ Z \rightarrow Z$, the application of $S_1$ for $\alpha=1$ leads to:
\begin{equation}
    \begin{split}
        & \rho''((-1)^1) =  \frac{1}{2}\sum_{i,k} D_{i,k} (I+(-1)^{\delta_{k,Z}}Y_1) \ , \\
        & D = \begin{pmatrix}
       \text{cos}^2(\pi/8) I & -(\frac{i}{2})\text{sin}(\pi/4) Z_1 \\ 
       (\frac{i}{2})\text{sin}(\pi/4) Z_1  & \text{sin}^2(\pi/8) I
    \end{pmatrix} \ .
    \end{split}
\end{equation}
This is just the state $R_Z(\theta=-\pi/4)\ket{Y} =\ket{T}$. Although this is a simple two qubit problem, we have now confirmed the Aaronson-Gottesman extended stabiliser approach for the $T$ gate by teleportation (equation \ref{eq:T_teleport_stab_try}). We shall move onto the more complicated Li \cite{Li_2015}/Lao-Criger \cite{LaoInjection2022} magic state injection protocol onto the surface code next.

\section{Probabilistic $\ket{T}$ State injection via stabiliser measurements}
Li developed a method to injection a magic state onto the un-rotated surface code probabilistically \cite{Li_2015}. Lao and Criger generalised this scheme to inject the bare physical magic state qubit located at the center or corner of a rotated surface code \cite{LaoInjection2022}. Both these procedure are quite similar, starting from a physical magic state qubit at the corner of the surface code, one can output a logical (surface code encoded) magic state qubit. The sub-figures \ref{fig:injection_init} to \ref{fig:injection_postselect} show the injection protocol for a corner qubit injected on the rotated surface code. 

\begin{figure}[!h]
    \centering
    \subfloat[][\label{fig:injection_init}Input arrangements of physical qubits on a grid for the magic state injection scheme \cite{Li_2015,LaoInjection2022}.]{    
    \resizebox{0.28\linewidth}{!}{
    \begin{tikzpicture}[scale=.7,every node/.style={minimum size=1cm},on grid]
    \begin{scope}[yshift=0,every node/.append style={yslant=0,xslant=0},yslant=0,xslant=0,rotate=0]
        \fill[white,fill opacity=0.0] (0,0) rectangle (10,4);
        \node[fill=green!50,shape=circle,draw=black] (n4) at (2,18) {${\ket{\psi}}$};
        \node[fill=red!50,shape=circle,draw=black] (n3) at (2,14) {$\ket{+}$};
        \node[fill=red!50,shape=circle,draw=black] (n2) at (2,10) {$\ket{+}$};
        \node[fill=red!50,shape=circle,draw=black] (n1) at (2,6) {$\ket{+}$};
        \node[fill=red!50,shape=circle,draw=black] (n0) at (2,2) {$\ket{+}$};

        \node[fill=blue!50,shape=circle,draw=black] (n9) at (6,18) {$\ket{0}$};
        \node[fill=red!50,shape=circle,draw=black] (n8) at (6,14) {$\ket{+}$};
        \node[fill=red!50,shape=circle,draw=black] (n7) at (6,10) {$\ket{+}$};
        \node[fill=red!50,shape=circle,draw=black] (n6) at (6,6) {$\ket{+}$};
        \node[fill=red!50,shape=circle,draw=black] (n5) at (6,2) {$\ket{+}$}; 

        \node[fill=blue!50,shape=circle,draw=black] (n14) at (10,18) {$\ket{0}$};
        \node[fill=blue!50,shape=circle,draw=black] (n13) at (10,14) {$\ket{0}$};
        \node[fill=red!50,shape=circle,draw=black] (n12) at (10,10) {$\ket{+}$};
        \node[fill=red!50,shape=circle,draw=black] (n11) at (10,6) {$\ket{+}$};
        \node[fill=red!50,shape=circle,draw=black] (n10) at (10,2) {$\ket{+}$};

        \node[fill=blue!50,shape=circle,draw=black] (n19) at (14,18) {$\ket{0}$};
        \node[fill=blue!50,shape=circle,draw=black] (n18) at (14,14) {$\ket{0}$};
        \node[fill=blue!50,shape=circle,draw=black] (n17) at (14,10) {$\ket{0}$};
        \node[fill=red!50,shape=circle,draw=black] (n16) at (14,6) {$\ket{+}$};
        \node[fill=red!50,shape=circle,draw=black] (n15) at (14,2) {$\ket{+}$}; 

        \node[fill=blue!50,shape=circle,draw=black] (n24) at (18,18) {$\ket{0}$};
        \node[fill=blue!50,shape=circle,draw=black] (n23) at (18,14) {$\ket{0}$};
        \node[fill=blue!50,shape=circle,draw=black] (n22) at (18,10) {$\ket{0}$};
        \node[fill=blue!50,shape=circle,draw=black] (n21) at (18,6) {$\ket{0}$};
        \node[fill=red!50,shape=circle,draw=black] (n20) at (18,2) {$\ket{+}$};

    \end{scope}

\end{tikzpicture}
}
    }
    \quad
    \subfloat[][\label{fig:injection_measure_stab}Measure all the $X$-type yellow and $Z$-type gray plaquettes for two full rounds of parity measurements.]{
    \resizebox{0.3\linewidth}{!}{
    \begin{tikzpicture}[scale=.7,every node/.style={minimum size=1cm},on grid]
    \begin{scope}[yshift=0,every node/.append style={yslant=0,xslant=0},yslant=0,xslant=0,rotate=0]
        \fill[white,fill opacity=0.0] (0,0) rectangle (10,4);
        \node[fill=green!50,shape=circle,draw=black] (n4) at (2,18) {${\ket{\psi}}$};
        \node[fill=red!50,shape=circle,draw=black] (n3) at (2,14) {$\ket{+}$};
        \node[fill=red!50,shape=circle,draw=black] (n2) at (2,10) {$\ket{+}$};
        \node[fill=red!50,shape=circle,draw=black] (n1) at (2,6) {$\ket{+}$};
        \node[fill=red!50,shape=circle,draw=black] (n0) at (2,2) {$\ket{+}$};

        \node[fill=blue!50,shape=circle,draw=black] (n9) at (6,18) {$\ket{0}$};
        \node[fill=red!50,shape=circle,draw=black] (n8) at (6,14) {$\ket{+}$};
        \node[fill=red!50,shape=circle,draw=black] (n7) at (6,10) {$\ket{+}$};
        \node[fill=red!50,shape=circle,draw=black] (n6) at (6,6) {$\ket{+}$};
        \node[fill=red!50,shape=circle,draw=black] (n5) at (6,2) {$\ket{+}$}; 

        \node[fill=blue!50,shape=circle,draw=black] (n14) at (10,18) {$\ket{0}$};
        \node[fill=blue!50,shape=circle,draw=black] (n13) at (10,14) {$\ket{0}$};
        \node[fill=red!50,shape=circle,draw=black] (n12) at (10,10) {$\ket{+}$};
        \node[fill=red!50,shape=circle,draw=black] (n11) at (10,6) {$\ket{+}$};
        \node[fill=red!50,shape=circle,draw=black] (n10) at (10,2) {$\ket{+}$};

        \node[fill=blue!50,shape=circle,draw=black] (n19) at (14,18) {$\ket{0}$};
        \node[fill=blue!50,shape=circle,draw=black] (n18) at (14,14) {$\ket{0}$};
        \node[fill=blue!50,shape=circle,draw=black] (n17) at (14,10) {$\ket{0}$};
        \node[fill=red!50,shape=circle,draw=black] (n16) at (14,6) {$\ket{+}$};
        \node[fill=red!50,shape=circle,draw=black] (n15) at (14,2) {$\ket{+}$}; 

        \node[fill=blue!50,shape=circle,draw=black] (n24) at (18,18) {$\ket{0}$};
        \node[fill=blue!50,shape=circle,draw=black] (n23) at (18,14) {$\ket{0}$};
        \node[fill=blue!50,shape=circle,draw=black] (n22) at (18,10) {$\ket{0}$};
        \node[fill=blue!50,shape=circle,draw=black] (n21) at (18,6) {$\ket{0}$};
        \node[fill=red!50,shape=circle,draw=black] (n20) at (18,2) {$\ket{+}$};

        \begin{pgfonlayer}{background}

        \draw[gray,fill=gray,opacity=0.65](1.8,2.2) to[curve through={(0,4)}](1.8,5.8);
        \filldraw[fill=yellow, opacity=0.65, draw=yellow] (2.2,2.2) rectangle (5.8,5.8);
        \filldraw[fill=gray, opacity=0.65, draw=gray] (2.2,6.2) rectangle (5.8,9.8);
        \filldraw[fill=yellow, opacity=0.65, draw=yellow] (2.2,10.2) rectangle (5.8,13.8);
        \draw[gray,fill=gray,opacity=0.65](1.8,10.2) to[curve through={(0,12)}](1.8,13.8);
        \filldraw[fill=gray, opacity=0.65, draw=gray] (2.2,14.2) rectangle (5.8,17.8);
        \draw[yellow,fill=yellow,opacity=0.65](2.2,18.2) to[curve through={(4,20)}](5.8,18.2);

        \draw[yellow,fill=yellow,opacity=0.65](9.8,1.8) to[curve through={(8,0)}](6.2,1.8);
        \filldraw[fill=gray, opacity=0.65, draw=gray] (6.2,2.2) rectangle (9.8,5.8);
        \filldraw[fill=yellow, opacity=0.65, draw=yellow] (6.2,6.2) rectangle (9.8,9.8);
        \filldraw[fill=gray, opacity=0.65, draw=gray] (6.2,10.2) rectangle (9.8,13.8);
        \filldraw[fill=yellow, opacity=0.65, draw=yellow] (6.2,14.2) rectangle (9.8,17.8);

        \filldraw[fill=yellow, opacity=0.65, draw=yellow] (10.2,2.2) rectangle (13.8,5.8);
        \filldraw[fill=gray, opacity=0.65, draw=gray] (10.2,6.2) rectangle (13.8,9.8);
        \filldraw[fill=yellow, opacity=0.65, draw=yellow] (10.2,10.2) rectangle (13.8,13.8);
        \filldraw[fill=gray, opacity=0.65, draw=gray] (10.2,14.2) rectangle (13.8,17.8);
        \draw[yellow,fill=yellow,opacity=0.65](10.2,18.2) to[curve through={(12,20)}](13.8,18.2);
        
        \draw[yellow,fill=yellow,opacity=0.65](17.8,1.8) to[curve through={(16,0)}](14.2,1.8);
        \filldraw[fill=gray, opacity=0.65, draw=gray] (14.2,2.2) rectangle (17.8,5.8);
        \filldraw[fill=yellow, opacity=0.65, draw=yellow] (14.2,6.2) rectangle (17.8,9.8);
        \draw[gray,fill=gray,opacity=0.65](18.2,9.8) to[curve through={(20,8)}](18.2,6.2);
        \filldraw[fill=gray, opacity=0.65, draw=gray] (14.2,10.2) rectangle (17.8,13.8);
        \filldraw[fill=yellow, opacity=0.65, draw=yellow] (14.2,14.2) rectangle (17.8,17.8);
        \draw[gray,fill=gray,opacity=0.65](18.2,17.8) to[curve through={(20,16)}](18.2,14.2);
        \end{pgfonlayer}
        
        \begin{pgfonlayer}{background}
            \path [-,line width=0.1cm,black,opacity=1] (n0) edge node {} (n20);
            \path [-,line width=0.1cm,black,opacity=1] (n1) edge node {} (n21);
            \path [-,line width=0.1cm,black,opacity=1] (n2) edge node {} (n22);
            \path [-,line width=0.1cm,black,opacity=1] (n3) edge node {} (n23);
            \path [-,line width=0.1cm,black,opacity=1] (n4) edge node {} (n24);

            \path [-,line width=0.1cm,black,opacity=1] (n0) edge node {} (n4);
            \path [-,line width=0.1cm,black,opacity=1] (n5) edge node {} (n9);
            \path [-,line width=0.1cm,black,opacity=1] (n10) edge node {} (n14);
            \path [-,line width=0.1cm,black,opacity=1] (n15) edge node {} (n19);
            \path [-,line width=0.1cm,black,opacity=1] (n20) edge node {} (n24);

            \node[fill=black,circle,draw=black,scale = 0.25] (b0) at (4,4) {};
            \node[fill=black,circle,draw=black,scale = 0.25] (b1) at (4,12) {};
            \node[fill=black,circle,draw=black,scale = 0.25] (b2) at (4,20) {};

            \node[fill=black,circle,draw=black,scale = 0.25] (b3) at (8,0) {};
            \node[fill=black,circle,draw=black,scale = 0.25] (b4) at (8,8) {};
            \node[fill=black,circle,draw=black,scale = 0.25] (b5) at (8,16) {};

            \node[fill=black,circle,draw=black,scale = 0.25] (b6) at (12,4) {};
            \node[fill=black,circle,draw=black,scale = 0.25] (b7) at (12,12) {};
            \node[fill=black,circle,draw=black,scale = 0.25] (b8) at (12,20) {};

            \node[fill=black,circle,draw=black,scale = 0.25] (b9) at (16,0) {};
            \node[fill=black,circle,draw=black,scale = 0.25] (b10) at (16,8) {};
            \node[fill=black,circle,draw=black,scale = 0.25] (b11) at (16,16) {};

            \node[fill=white,circle,draw=black,scale = 0.25] (a0) at (0,4) {};
            \node[fill=white,circle,draw=black,scale = 0.25] (a1) at (0,12) {};

            \node[fill=white,circle,draw=black,scale = 0.25] (a2) at (4,8) {};
            \node[fill=white,circle,draw=black,scale = 0.25] (a3) at (4,16) {};

            \node[fill=white,circle,draw=black,scale = 0.25] (a4) at (8,4) {};
            \node[fill=white,circle,draw=black,scale = 0.25] (a5) at (8,12) {};

            \node[fill=white,circle,draw=black,scale = 0.25] (a6) at (12,8) {};
            \node[fill=white,circle,draw=black,scale = 0.25] (a7) at (12,16) {};

            \node[fill=white,circle,draw=black,scale = 0.25] (a8) at (16,4) {};
            \node[fill=white,circle,draw=black,scale = 0.25] (a9) at (16,12) {};

            \node[fill=white,circle,draw=black,scale = 0.25] (a10) at (20,8) {};
            \node[fill=white,circle,draw=black,scale = 0.25] (a11) at (20,16) {};
    
        \end{pgfonlayer}
    \end{scope}

\end{tikzpicture}
}
    }
\quad
\subfloat[][\label{fig:injection_postselect}After the parity measurements in figure \ref{fig:injection_measure_stab}, post-select upon all $+1$ on all coloured plaquettes.]{  
\resizebox{0.30\linewidth}{!}{
    \begin{tikzpicture}[scale=.7,every node/.style={minimum size=1cm},on grid]
    \begin{scope}[yshift=0,every node/.append style={yslant=0,xslant=0},yslant=0,xslant=0,rotate=0]
        \fill[white,fill opacity=0.0] (0,0) rectangle (10,4);
        \node[fill=green!50,shape=circle,draw=black] (n4) at (2,18) {${\ket{\psi}}$};
        \node[fill=red!50,shape=circle,draw=black] (n3) at (2,14) {$\ket{+}$};
        \node[fill=red!50,shape=circle,draw=black] (n2) at (2,10) {$\ket{+}$};
        \node[fill=red!50,shape=circle,draw=black] (n1) at (2,6) {$\ket{+}$};
        \node[fill=red!50,shape=circle,draw=black] (n0) at (2,2) {$\ket{+}$};

        \node[fill=blue!50,shape=circle,draw=black] (n9) at (6,18) {$\ket{0}$};
        \node[fill=red!50,shape=circle,draw=black] (n8) at (6,14) {$\ket{+}$};
        \node[fill=red!50,shape=circle,draw=black] (n7) at (6,10) {$\ket{+}$};
        \node[fill=red!50,shape=circle,draw=black] (n6) at (6,6) {$\ket{+}$};
        \node[fill=red!50,shape=circle,draw=black] (n5) at (6,2) {$\ket{+}$}; 

        \node[fill=blue!50,shape=circle,draw=black] (n14) at (10,18) {$\ket{0}$};
        \node[fill=blue!50,shape=circle,draw=black] (n13) at (10,14) {$\ket{0}$};
        \node[fill=red!50,shape=circle,draw=black] (n12) at (10,10) {$\ket{+}$};
        \node[fill=red!50,shape=circle,draw=black] (n11) at (10,6) {$\ket{+}$};
        \node[fill=red!50,shape=circle,draw=black] (n10) at (10,2) {$\ket{+}$};

        \node[fill=blue!50,shape=circle,draw=black] (n19) at (14,18) {$\ket{0}$};
        \node[fill=blue!50,shape=circle,draw=black] (n18) at (14,14) {$\ket{0}$};
        \node[fill=blue!50,shape=circle,draw=black] (n17) at (14,10) {$\ket{0}$};
        \node[fill=red!50,shape=circle,draw=black] (n16) at (14,6) {$\ket{+}$};
        \node[fill=red!50,shape=circle,draw=black] (n15) at (14,2) {$\ket{+}$}; 

        \node[fill=blue!50,shape=circle,draw=black] (n24) at (18,18) {$\ket{0}$};
        \node[fill=blue!50,shape=circle,draw=black] (n23) at (18,14) {$\ket{0}$};
        \node[fill=blue!50,shape=circle,draw=black] (n22) at (18,10) {$\ket{0}$};
        \node[fill=blue!50,shape=circle,draw=black] (n21) at (18,6) {$\ket{0}$};
        \node[fill=red!50,shape=circle,draw=black] (n20) at (18,2) {$\ket{+}$};

        \begin{pgfonlayer}{background}

        \filldraw[fill=yellow, opacity=0.65, draw=yellow] (2.2,2.2) rectangle (5.8,5.8);
        \filldraw[fill=yellow, opacity=0.65, draw=yellow] (2.2,10.2) rectangle (5.8,13.8);

        \draw[yellow,fill=yellow,opacity=0.65](9.8,1.8) to[curve through={(8,0)}](6.2,1.8);
        \filldraw[fill=yellow, opacity=0.65, draw=yellow] (6.2,6.2) rectangle (9.8,9.8);

        \filldraw[fill=yellow, opacity=0.65, draw=yellow] (10.2,2.2) rectangle (13.8,5.8);
        \filldraw[fill=gray, opacity=0.65, draw=gray] (10.2,14.2) rectangle (13.8,17.8);
        
        \draw[yellow,fill=yellow,opacity=0.65](17.8,1.8) to[curve through={(16,0)}](14.2,1.8);
        \draw[gray,fill=gray,opacity=0.65](18.2,9.8) to[curve through={(20,8)}](18.2,6.2);
        \filldraw[fill=gray, opacity=0.65, draw=gray] (14.2,10.2) rectangle (17.8,13.8);
        \draw[gray,fill=gray,opacity=0.65](18.2,17.8) to[curve through={(20,16)}](18.2,14.2);
        \end{pgfonlayer}
        
        \begin{pgfonlayer}{background}
            \path [-,line width=0.1cm,black,opacity=1] (n0) edge node {} (n20);
            \path [-,line width=0.1cm,black,opacity=1] (n1) edge node {} (n21);
            \path [-,line width=0.1cm,black,opacity=1] (n2) edge node {} (n22);
            \path [-,line width=0.1cm,black,opacity=1] (n3) edge node {} (n23);
            \path [-,line width=0.1cm,black,opacity=1] (n4) edge node {} (n24);

            \path [-,line width=0.1cm,black,opacity=1] (n0) edge node {} (n4);
            \path [-,line width=0.1cm,black,opacity=1] (n5) edge node {} (n9);
            \path [-,line width=0.1cm,black,opacity=1] (n10) edge node {} (n14);
            \path [-,line width=0.1cm,black,opacity=1] (n15) edge node {} (n19);
            \path [-,line width=0.1cm,black,opacity=1] (n20) edge node {} (n24);

            \node[fill=black,circle,draw=black,scale = 0.25] (b0) at (4,4) {};
            \node[fill=black,circle,draw=black,scale = 0.25] (b1) at (4,12) {};
            \node[fill=black,circle,draw=black,scale = 0.25] (b2) at (4,20) {};

            \node[fill=black,circle,draw=black,scale = 0.25] (b3) at (8,0) {};
            \node[fill=black,circle,draw=black,scale = 0.25] (b4) at (8,8) {};
            \node[fill=black,circle,draw=black,scale = 0.25] (b5) at (8,16) {};

            \node[fill=black,circle,draw=black,scale = 0.25] (b6) at (12,4) {};
            \node[fill=black,circle,draw=black,scale = 0.25] (b7) at (12,12) {};
            \node[fill=black,circle,draw=black,scale = 0.25] (b8) at (12,20) {};

            \node[fill=black,circle,draw=black,scale = 0.25] (b9) at (16,0) {};
            \node[fill=black,circle,draw=black,scale = 0.25] (b10) at (16,8) {};
            \node[fill=black,circle,draw=black,scale = 0.25] (b11) at (16,16) {};

            \node[fill=white,circle,draw=black,scale = 0.25] (a0) at (0,4) {};
            \node[fill=white,circle,draw=black,scale = 0.25] (a1) at (0,12) {};

            \node[fill=white,circle,draw=black,scale = 0.25] (a2) at (4,8) {};
            \node[fill=white,circle,draw=black,scale = 0.25] (a3) at (4,16) {};

            \node[fill=white,circle,draw=black,scale = 0.25] (a4) at (8,4) {};
            \node[fill=white,circle,draw=black,scale = 0.25] (a5) at (8,12) {};

            \node[fill=white,circle,draw=black,scale = 0.25] (a6) at (12,8) {};
            \node[fill=white,circle,draw=black,scale = 0.25] (a7) at (12,16) {};

            \node[fill=white,circle,draw=black,scale = 0.25] (a8) at (16,4) {};
            \node[fill=white,circle,draw=black,scale = 0.25] (a9) at (16,12) {};

            \node[fill=white,circle,draw=black,scale = 0.25] (a10) at (20,8) {};
            \node[fill=white,circle,draw=black,scale = 0.25] (a11) at (20,16) {};
    
        \end{pgfonlayer}
    \end{scope}

\end{tikzpicture}
}
}

\caption{}
\end{figure}

We start with this arrangements of physical qubits on a grid. Top left corner qubit initialised in the $\ket{\psi}$ (green node) state, $\ket{+}$ along the lower triangular + diagonals (red node), and $\ket{0}$ along the upper triangular (blue node) (figure \ref{fig:injection_init}). After the initialisation step in figure \ref {fig:injection_init}, measure all the stabilisers for 2 rounds of parity measurements. The yellow plaquettes are the $X$-type stabiliser while the gray plaquettes are the $Z$-type stabiliser of the rotated surface code (as shown in figure \ref{fig:injection_measure_stab}). After the two full rounds parity measurements, post-select on all the restricted set of coloured plaquette in figure \ref{fig:injection_postselect} returning $+1$ parity value. If any of those parity measurements returns $-1$, restart at the step outlined in figure \ref{fig:injection_init}). If all the parity measurements are $+1$ for both rounds of stabiliser measurements, you can proceed to grow the state \cite{Li_2015,LaoInjection2022} or end the procedure by performing one more additional rounds of stabiliser measurements, then you have encoded a logical $\ket{\bar{\psi}}$. We shall ignore the distance growth procedure now, please refer to \cite{Li_2015,LaoInjection2022} for discussion or \cite{wan2025pauliwebyranglestate} for a ZX calculus flavoured investigation of the probabilistic state injection problem.

\section{Magic state injection on $[[4,1,2]]$}
We will try to perform the magic state injection on the simpler $d=2$ rotated surface code as a toy problem first.
\begin{figure}[!h]
\centering\caption{\label{fig:injection_labelled_qb_4_1_2}The $[[4,1,2]]$ surface code with yellow = $XXXX$ plaquette and grey = $ZZ$ plaquettes, the qubits coloured {\color{red}red}, {\color{green}green} and {\color{blue}blue} are initialised in the {\color{red}$\ket{+}$}, {\color{green}$\ket{T}$} and {\color{blue}$\ket{0}$} states respectively.}
\resizebox{0.4\linewidth}{!}{
    \begin{tikzpicture}[scale=.7,every node/.style={minimum size=1cm},on grid]
    \begin{scope}[yshift=0,every node/.append style={yslant=0,xslant=0},yslant=0,xslant=0,rotate=0]
        \fill[white,fill opacity=0.0] (0,0) rectangle (8,4);

        \node[fill=green!50,shape=circle,draw=black] (n1) at (2,6) {$1$};
        \node[fill=red!50,shape=circle,draw=black] (n0) at (2,2) {$0$};

        \node[fill=blue!50,shape=circle,draw=black] (n6) at (6,6) {$3$};
        \node[fill=red!50,shape=circle,draw=black] (n5) at (6,2) {$2$};

        \begin{pgfonlayer}{background}

        \draw[gray,fill=gray,opacity=0.65](1.8,2.2) to[curve through={(0,4)}](1.8,5.8);
        \draw[gray,fill=gray,opacity=0.65](6.2,2.2) to[curve through={(8,4)}](6.2,5.8);
        \filldraw[fill=yellow, opacity=0.65, draw=yellow] (2.2,2.2) rectangle (5.8,5.8);
        
        \path [-,line width=0.1cm,black,opacity=1] (n0) edge node {} (n1);
        \path [-,line width=0.1cm,black,opacity=1] (n1) edge node {} (n6);
        \path [-,line width=0.1cm,black,opacity=1] (n6) edge node {} (n5);
        \path [-,line width=0.1cm,black,opacity=1] (n5) edge node {} (n0);

        \node[fill=black,circle,draw=black,scale = 0.25] (b0) at (4,4) {};
            
        \node[fill=white,circle,draw=black,scale = 0.25] (a0) at (0,4) {};
        
        \node[fill=white,circle,draw=black,scale = 0.25] (a1) at (8,4) {};
        
        \end{pgfonlayer}
    \end{scope}

\end{tikzpicture}
}
\end{figure}
Figure \ref{fig:injection_labelled_qb_4_1_2} shows the $[[4,1,2]]$ error detecting code. We wish to inject a magic state on this with Lao-Criger's scheme \cite{LaoInjection2022} as a toy example. The full $\ket{T}$ state injection circuit is given by equation \ref{eq:4_1_2_init_circuit}, 
\begin{equation}
\label{eq:4_1_2_init_circuit}
\centering
\resizebox{0.95\linewidth}{!}{%
    \begin{tikzpicture}
    \begin{yquant}
    qubit {$\ket{+}_0$} q[+1];
    qubit {$\ket{+}_1$} q[+1];
    qubit {$\ket{+}_2$} q[+1];
    qubit {$\ket{0}_3$} q[+1];
    [blue, thick, label={\small$\tau = 0$}]    barrier (-);
    box {$T$} q[1];
    [blue, thick, label={\small$\tau = 1$}]    barrier (-);
    align  q;
    box {$M_{ZZ} = n_0$} (q[0]-q[1]);
    box {$M_{ZZ} = n_1$} (q[2]-q[3]);
    [blue, thick, label={\small$\tau = 2$}]    barrier (-);
    align  q;
    box {$M_{XXXX} = m$} (q[0]-q[3]);
    [blue, thick, label={\small$\tau = 3$}]    barrier (-);
    \end{yquant}
    \end{tikzpicture} \ ,
    }
\end{equation}
where $M_{ZZ}$ and $M_{XXXX}$ are parity measurements with their respective measurement values $n_0,n_1$ and $m$. Let's see if this works. 
\subsection{$[[4,1,2]]$ code stabilisers and logical operators}
Let's remind ourselves what the stabiliser and logical operators of the $[[4,1,2]]$ code is \cite{eczoo_bacon_shor_4}:
\begin{equation}
    \label{eq:4_1_2_stab_log}
    \begin{split}
        & \langle Z_0Z_1, Z_2Z_3, X_0X_1X_2X_3 \rangle \\ & \bar{Z} = Z_1Z_3 \ , \ \bar{X} = X_0 X_1 \ .
    \end{split}
\end{equation}
\subsection{$M_{ZZ}$ measurements}
We first perform the two-body $Z$-type stabiliser parity measurements.
\subsubsection{$M_{Z_2Z_3}$ measurement}
The state is after $\tau = 1$:
\begin{equation}
    \begin{split}
        \rho' & =  \frac{1}{2^4}\sum_{i,k} D_{i,k} (I+(-1)^{\omega(X_1, P_k)}X_1) \\ & \ \ \ \ \ \ \cdot (I+X_0)(I+X_2)(I+Z_3) \ , \\ 
        D & = \begin{pmatrix}
       \text{cos}^2(\pi/8) I & (i/2)\text{sin}(\pi/4) Z_1 \\ 
       -(i/2)\text{sin}(\pi/4) Z_1  & \text{sin}^2(\pi/8) I
    \end{pmatrix} \ .
    \end{split}
\end{equation}
Suppose we perform a $M_{ZZ}$ measurement on qubits $2$ and $3$, with measurement result $n_1$:
\begin{equation}
    (I+X_2)(I+Z_3) \rightarrow (I+Z_3)(I+(-1)^{n_1}Z_2Z_3) \ .
\end{equation}
This implies:
\begin{equation}
    \begin{split}
        \rho' & =  \frac{1}{2^4}\sum_{i,k} D_{i,k} (I+(-1)^{\omega(X_1, P_k)}X_1) \cdot (I+X_0)(I+Z_3)(I+(-1)^{n_1}Z_2Z_3) \ , \\ 
        D & = \begin{pmatrix}
       \text{cos}^2(\pi/8) I & (i/2)\text{sin}(\pi/4) Z_1 \\ 
       -(i/2)\text{sin}(\pi/4) Z_1  & \text{sin}^2(\pi/8) I
    \end{pmatrix} \ .
    \end{split}
\end{equation}

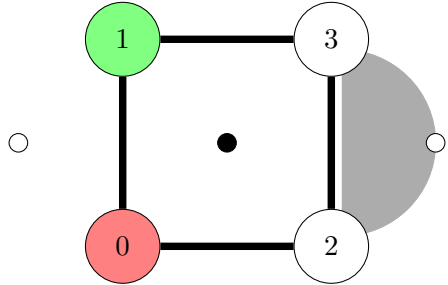
\begin{figure}[!h]
\centering\caption{\label{fig:injection_labelled_qb_4_1_2_Z2Z3_meas}After the $M_{Z_2Z_3}$ measurement.}
\resizebox{0.4\linewidth}{!}{
    \begin{tikzpicture}[scale=.7,every node/.style={minimum size=1cm},on grid]
    \begin{scope}[yshift=0,every node/.append style={yslant=0,xslant=0},yslant=0,xslant=0,rotate=0]
        \fill[white,fill opacity=0.0] (0,0) rectangle (8,4);

        \node[fill=green!50,shape=circle,draw=black] (n1) at (2,6) {$1$};
        \node[fill=red!50,shape=circle,draw=black] (n0) at (2,2) {$0$};

        \node[fill=white!50,shape=circle,draw=black] (n6) at (6,6) {$3$};
        \node[fill=white!50,shape=circle,draw=black] (n5) at (6,2) {$2$};

        \begin{pgfonlayer}{background}

        \draw[gray,fill=gray,opacity=0.65](6.2,2.2) to[curve through={(8,4)}](6.2,5.8);
        
        \path [-,line width=0.1cm,black,opacity=1] (n0) edge node {} (n1);
        \path [-,line width=0.1cm,black,opacity=1] (n1) edge node {} (n6);
        \path [-,line width=0.1cm,black,opacity=1] (n6) edge node {} (n5);
        \path [-,line width=0.1cm,black,opacity=1] (n5) edge node {} (n0);

        \node[fill=black,circle,draw=black,scale = 0.25] (b0) at (4,4) {};
            
        \node[fill=white,circle,draw=black,scale = 0.25] (a0) at (0,4) {};
        
        \node[fill=white,circle,draw=black,scale = 0.25] (a1) at (8,4) {};
        
        \end{pgfonlayer}
    \end{scope}

\end{tikzpicture}
}
\end{figure}

\subsubsection{$M_{Z_0Z_1}$ measurement}

For the $M_{ZZ}$ measurement on qubits $0$ and $1$, let's re-write stabilisers:
\begin{equation}
\begin{split}
        & (I+(-1)^{\omega(X_1, P_k)}X_1)(I+X_0) \rightarrow (I+(-1)^{\omega(X_1, P_k)}X_1)(I+(-1)^{\omega(X_1, P_k)}X_0X_1) \ .
        \end{split}
\end{equation}
If we were to measure in the $Q = Z_0Z_1$ basis now:
\begin{equation}
    \begin{split}
        \rho' & \propto \sum_{i,k} D_{i,k} (I+(-1)^{n_0}Z_0Z_1) \cdot (I+(-1)^{\omega(X_1, P_k)}X_0X_1) \cdot (I+Z_3)(I+(-1)^{n_1}Z_2Z_3) \ , \\ 
        D & = \begin{pmatrix}
       \text{cos}^2(\pi/8) I & (i/2)\text{sin}(\pi/4) Z_1 \\ 
       -(i/2)\text{sin}(\pi/4) Z_1  & \text{sin}^2(\pi/8) I
    \end{pmatrix} \ .
    \end{split}
\end{equation}
Note that $(I+(-1)^{n_0}Z_0Z_1)$ and $(I+(-1)^{\omega(X_1, P_k)}X_0X_1)$ commutes, so you can push it through.
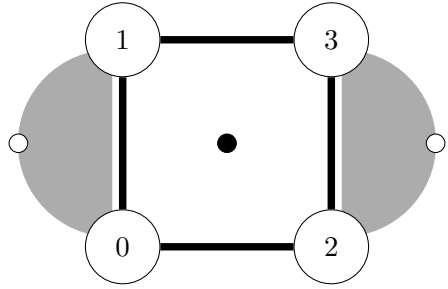
\begin{figure}[!h]
\centering\caption{\label{fig:injection_labelled_qb_4_1_2_Z0Z1_meas}After the $M_{Z_0Z_1}$ measurement.}
\resizebox{0.4\linewidth}{!}{
    \begin{tikzpicture}[scale=.7,every node/.style={minimum size=1cm},on grid]
    \begin{scope}[yshift=0,every node/.append style={yslant=0,xslant=0},yslant=0,xslant=0,rotate=0]
        \fill[white,fill opacity=0.0] (0,0) rectangle (8,4);

        \node[fill=white!50,shape=circle,draw=black] (n1) at (2,6) {$1$};
        \node[fill=white!50,shape=circle,draw=black] (n0) at (2,2) {$0$};

        \node[fill=white!50,shape=circle,draw=black] (n6) at (6,6) {$3$};
        \node[fill=white!50,shape=circle,draw=black] (n5) at (6,2) {$2$};

        \begin{pgfonlayer}{background}

        \draw[gray,fill=gray,opacity=0.65](1.8,2.2) to[curve through={(0,4)}](1.8,5.8);
        \draw[gray,fill=gray,opacity=0.65](6.2,2.2) to[curve through={(8,4)}](6.2,5.8);
        
        \path [-,line width=0.1cm,black,opacity=1] (n0) edge node {} (n1);
        \path [-,line width=0.1cm,black,opacity=1] (n1) edge node {} (n6);
        \path [-,line width=0.1cm,black,opacity=1] (n6) edge node {} (n5);
        \path [-,line width=0.1cm,black,opacity=1] (n5) edge node {} (n0);

        \node[fill=black,circle,draw=black,scale = 0.25] (b0) at (4,4) {};
            
        \node[fill=white,circle,draw=black,scale = 0.25] (a0) at (0,4) {};
        
        \node[fill=white,circle,draw=black,scale = 0.25] (a1) at (8,4) {};
        
        \end{pgfonlayer}
    \end{scope}

\end{tikzpicture}
}
\end{figure}
\subsection{$M_{XXXX}$ measurement}
We now measure in the $Q = X_0X_1X_2X_3$ basis, with measurement results $m$. Note that the off-diagonal $D_{i,k}$ anti-commutes with the $X_0X_1X_2X_3$ and $X_0X_1X_2X_3$ also anti-commutes with $Z_3$. Applying the update rule 2 (equation \ref{eq:update_rule_2}).
\begin{equation}
    \begin{split}
        \rho' & \propto \sum_{i,k} D_{i,k} (I+(-1)^{\omega(X_1, P_k)}X_0X_1) \cdot (I+(-1)^{n_0}Z_0Z_1)(I+(-1)^{n_1}Z_2Z_3)  \\ & \ \ \ \ \ \ \cdot (I+(-1)^mX_0X_1X_2X_3) \ , \\ 
        D & = \begin{pmatrix}
       \text{cos}^2(\pi/8) I & (i/2)\text{sin}(\pi/4) Z_1Z_3 \\ 
       -(i/2)\text{sin}(\pi/4) Z_1Z_3  & \text{sin}^2(\pi/8) I
    \end{pmatrix} \ .
    \end{split}
\end{equation}
Now, re-write $X_0X_1 = \bar{X}$ and $Z_1Z_3 = \bar{Z}$, the logical operators of the $[[4,1,2]]$ code (equation \ref{eq:4_1_2_stab_log}). This implies:
\begin{equation}
    \begin{split}
        \rho' & \propto \sum_{i,k} D_{i,k} (I+(-1)^{\omega(X_1, P_k)}\bar{X}) \cdot (I+(-1)^{n_0}Z_0Z_1)(I+(-1)^{n_1}Z_2Z_3)  \\ & \ \ \ \ \ \ \cdot (I+(-1)^mX_0X_1X_2X_3) \ , \\ 
        D & = \begin{pmatrix}
       \text{cos}^2(\pi/8) I & (i/2)\text{sin}(\pi/4) \bar{Z} \\ 
       -(i/2)\text{sin}(\pi/4) \bar{Z}  & \text{sin}^2(\pi/8) I
    \end{pmatrix} \ .
    \end{split}
\end{equation}
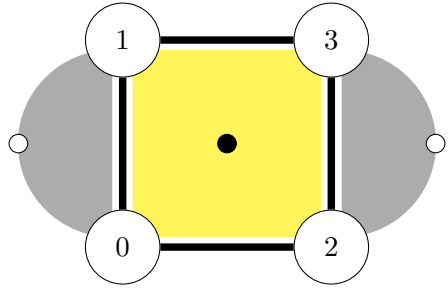
\begin{figure}[!h]
\centering\caption{\label{fig:injection_labelled_qb_4_1_2_XXXX_meas}After the $M_{X_0X_1X_2X_3}$ measurement.}
\resizebox{0.4\linewidth}{!}{
    \begin{tikzpicture}[scale=.7,every node/.style={minimum size=1cm},on grid]
    \begin{scope}[yshift=0,every node/.append style={yslant=0,xslant=0},yslant=0,xslant=0,rotate=0]
        \fill[white,fill opacity=0.0] (0,0) rectangle (8,4);

        \node[fill=white!50,shape=circle,draw=black] (n1) at (2,6) {$1$};
        \node[fill=white!50,shape=circle,draw=black] (n0) at (2,2) {$0$};

        \node[fill=white!50,shape=circle,draw=black] (n6) at (6,6) {$3$};
        \node[fill=white!50,shape=circle,draw=black] (n5) at (6,2) {$2$};

        \begin{pgfonlayer}{background}

        \draw[gray,fill=gray,opacity=0.65](1.8,2.2) to[curve through={(0,4)}](1.8,5.8);
        \draw[gray,fill=gray,opacity=0.65](6.2,2.2) to[curve through={(8,4)}](6.2,5.8);
        \filldraw[fill=yellow, opacity=0.65, draw=yellow] (2.2,2.2) rectangle (5.8,5.8);
        
        \path [-,line width=0.1cm,black,opacity=1] (n0) edge node {} (n1);
        \path [-,line width=0.1cm,black,opacity=1] (n1) edge node {} (n6);
        \path [-,line width=0.1cm,black,opacity=1] (n6) edge node {} (n5);
        \path [-,line width=0.1cm,black,opacity=1] (n5) edge node {} (n0);

        \node[fill=black,circle,draw=black,scale = 0.25] (b0) at (4,4) {};
            
        \node[fill=white,circle,draw=black,scale = 0.25] (a0) at (0,4) {};
        
        \node[fill=white,circle,draw=black,scale = 0.25] (a1) at (8,4) {};
        
        \end{pgfonlayer}
    \end{scope}

\end{tikzpicture}
}
\end{figure}

Note that $(-1)^{\omega(X_1,P_k)} = 1$ for the first column of $k$ and $(-1)^{\omega(X_1,P_k)} = -1$ for the second column of $k$. Hence, the state produced is the logical $\ket{\bar{T}}$ state with the correct stabilisers (Pauli-frame dependent on measurement values: $n_0,n_1,m$). This is the $d=2$ logical $\ket{\bar{T}}$ state, assuming no errors had occur during the injection process. 

\section{Intuition}
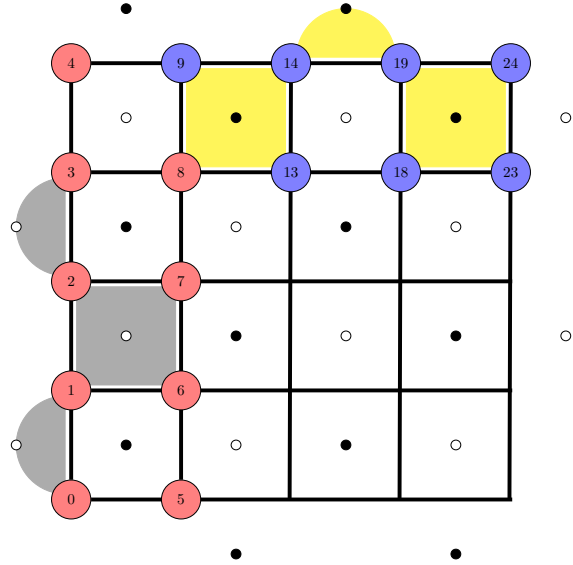
\begin{figure}[!h]
\centering
\caption{\label{fig:injection_labelled_qb_reduced}Qubits are labelled now, yellow = $XXXX$ plaquettes, grey = $ZZZZ$ plaquettes.}
\resizebox{0.5\linewidth}{!}{
    \begin{tikzpicture}[scale=.7,every node/.style={minimum size=1cm},on grid]
    \begin{scope}[yshift=0,every node/.append style={yslant=0,xslant=0},yslant=0,xslant=0,rotate=0]
        \fill[white,fill opacity=0.0] (0,0) rectangle (10,4);
        \node[fill=red!50,shape=circle,draw=black] (n4) at (2,18) {$4$};
        \node[fill=red!50,shape=circle,draw=black] (n3) at (2,14) {$3$};
        \node[fill=red!50,shape=circle,draw=black] (n2) at (2,10) {$2$};
        \node[fill=red!50,shape=circle,draw=black] (n1) at (2,6) {$1$};
        \node[fill=red!50,shape=circle,draw=black] (n0) at (2,2) {$0$};

        \node[fill=blue!50,shape=circle,draw=black] (n9) at (6,18) {$9$};
        \node[fill=red!50,shape=circle,draw=black] (n8) at (6,14) {$8$};
        \node[fill=red!50,shape=circle,draw=black] (n7) at (6,10) {$7$};
        \node[fill=red!50,shape=circle,draw=black] (n6) at (6,6) {$6$};
        \node[fill=red!50,shape=circle,draw=black] (n5) at (6,2) {$5$}; 

        \node[fill=blue!50,shape=circle,draw=black] (n14) at (10,18) {$14$};
        \node[fill=blue!50,shape=circle,draw=black] (n13) at (10,14) {$13$};

        \node[fill=blue!50,shape=circle,draw=black] (n19) at (14,18) {$19$};
        \node[fill=blue!50,shape=circle,draw=black] (n18) at (14,14) {$18$};

        \node[fill=blue!50,shape=circle,draw=black] (n24) at (18,18) {$24$};
        \node[fill=blue!50,shape=circle,draw=black] (n23) at (18,14) {$23$};

        \begin{pgfonlayer}{background}

        \draw[gray,fill=gray,opacity=0.65](1.8,2.2) to[curve through={(0,4)}](1.8,5.8);
        \filldraw[fill=gray, opacity=0.65, draw=gray] (2.2,6.2) rectangle (5.8,9.8);
        \draw[gray,fill=gray,opacity=0.65](1.8,10.2) to[curve through={(0,12)}](1.8,13.8);

        \filldraw[fill=yellow, opacity=0.65, draw=yellow] (6.2,14.2) rectangle (9.8,17.8);

        \draw[yellow,fill=yellow,opacity=0.65](10.2,18.2) to[curve through={(12,20)}](13.8,18.2);
        
        \filldraw[fill=yellow, opacity=0.65, draw=yellow] (14.2,14.2) rectangle (17.8,17.8);
        \end{pgfonlayer}
        
        \begin{pgfonlayer}{background}
            \path [-,line width=0.1cm,black,opacity=1] (n0) edge node {} ($(n20)+(.05,0)$);
            \path [-,line width=0.1cm,black,opacity=1] (n1) edge node {} ($(n21)+(.05,0)$);
            \path [-,line width=0.1cm,black,opacity=1] (n2) edge node {} ($(n22)+(.05,0)$);
            \path [-,line width=0.1cm,black,opacity=1] (n3) edge node {} (n23);
            \path [-,line width=0.1cm,black,opacity=1] (n4) edge node {} (n24);

            \path [-,line width=0.1cm,black,opacity=1] (n0) edge node {} (n4);
            \path [-,line width=0.1cm,black,opacity=1] (n5) edge node {} (n9);
            \path [-,line width=0.1cm,black,opacity=1] ($(n10)+(-.05,0)$) edge node {} (n14);
            \path [-,line width=0.1cm,black,opacity=1] ($(n15)+(-.05,0)$) edge node {} (n19);
            \path [-,line width=0.1cm,black,opacity=1] ($(n20)+(-.05,0)$) edge node {} (n24);

            \node[fill=black,circle,draw=black,scale = 0.25] (b0) at (4,4) {};
            \node[fill=black,circle,draw=black,scale = 0.25] (b1) at (4,12) {};
            \node[fill=black,circle,draw=black,scale = 0.25] (b2) at (4,20) {};

            \node[fill=black,circle,draw=black,scale = 0.25] (b3) at (8,0) {};
            \node[fill=black,circle,draw=black,scale = 0.25] (b4) at (8,8) {};
            \node[fill=black,circle,draw=black,scale = 0.25] (b5) at (8,16) {};

            \node[fill=black,circle,draw=black,scale = 0.25] (b6) at (12,4) {};
            \node[fill=black,circle,draw=black,scale = 0.25] (b7) at (12,12) {};
            \node[fill=black,circle,draw=black,scale = 0.25] (b8) at (12,20) {};

            \node[fill=black,circle,draw=black,scale = 0.25] (b9) at (16,0) {};
            \node[fill=black,circle,draw=black,scale = 0.25] (b10) at (16,8) {};
            \node[fill=black,circle,draw=black,scale = 0.25] (b11) at (16,16) {};

            \node[fill=white,circle,draw=black,scale = 0.25] (a0) at (0,4) {};
            \node[fill=white,circle,draw=black,scale = 0.25] (a1) at (0,12) {};

            \node[fill=white,circle,draw=black,scale = 0.25] (a2) at (4,8) {};
            \node[fill=white,circle,draw=black,scale = 0.25] (a3) at (4,16) {};

            \node[fill=white,circle,draw=black,scale = 0.25] (a4) at (8,4) {};
            \node[fill=white,circle,draw=black,scale = 0.25] (a5) at (8,12) {};

            \node[fill=white,circle,draw=black,scale = 0.25] (a6) at (12,8) {};
            \node[fill=white,circle,draw=black,scale = 0.25] (a7) at (12,16) {};

            \node[fill=white,circle,draw=black,scale = 0.25] (a8) at (16,4) {};
            \node[fill=white,circle,draw=black,scale = 0.25] (a9) at (16,12) {};

            \node[fill=white,circle,draw=black,scale = 0.25] (a10) at (20,8) {};
            \node[fill=white,circle,draw=black,scale = 0.25] (a11) at (20,16) {};
    
        \end{pgfonlayer}
    \end{scope}

\end{tikzpicture}
}
\end{figure}

The goal now is to study how to make this fault tolerant and post select upon the state given the measurement results. We shall ignore the details behind intricacies like patch distance growth or why $2$ full rounds of $Z$ then $X$ parity measurements are needed in Li's original work \cite{Li_2015}.
Let's look at the injection protocol on a larger $d=5$ rotated surface code in figure \ref{fig:injection_labelled_qb_reduced}.
We measure all the stabilisers that are coloured in figure \ref{fig:injection_labelled_qb_reduced} before initialising the $\ket{T}$ state in the corner by apply $T_4\ket{+}_4$.

\subsection{$Z$ stabiliser}
Suppose we measure with $Z$ stabiliser generators: $Z_0Z_1,Z_1Z_2Z_6Z_7,Z_3Z_4$ as shown in figure \ref{fig:injection_labelled_qb_reduced_Z}: 
\begin{figure}[!h]
\centering
\caption{\label{fig:injection_labelled_qb_reduced_Z}}
\resizebox{0.5\linewidth}{!}{
    \begin{tikzpicture}[scale=.7,every node/.style={minimum size=1cm},on grid]
    \begin{scope}[yshift=0,every node/.append style={yslant=0,xslant=0},yslant=0,xslant=0,rotate=0]
        \fill[white,fill opacity=0.0] (0,0) rectangle (10,4);
        \node[fill=red!50,shape=circle,draw=black] (n4) at (2,18) {$4$};
        \node[fill=red!50,shape=circle,draw=black] (n3) at (2,14) {$3$};
        \node[fill=red!50,shape=circle,draw=black] (n2) at (2,10) {$2$};
        \node[fill=red!50,shape=circle,draw=black] (n1) at (2,6) {$1$};
        \node[fill=red!50,shape=circle,draw=black] (n0) at (2,2) {$0$};

        \node[fill=blue!50,shape=circle,draw=black] (n9) at (6,18) {$9$};
        \node[fill=red!50,shape=circle,draw=black] (n8) at (6,14) {$8$};
        \node[fill=red!50,shape=circle,draw=black] (n7) at (6,10) {$7$};
        \node[fill=red!50,shape=circle,draw=black] (n6) at (6,6) {$6$};
        \node[fill=red!50,shape=circle,draw=black] (n5) at (6,2) {$5$}; 

        \node[fill=blue!50,shape=circle,draw=black] (n14) at (10,18) {$14$};
        \node[fill=blue!50,shape=circle,draw=black] (n13) at (10,14) {$13$};

        \node[fill=blue!50,shape=circle,draw=black] (n19) at (14,18) {$19$};
        \node[fill=blue!50,shape=circle,draw=black] (n18) at (14,14) {$18$};

        \node[fill=blue!50,shape=circle,draw=black] (n24) at (18,18) {$24$};
        \node[fill=blue!50,shape=circle,draw=black] (n23) at (18,14) {$23$};

        \begin{pgfonlayer}{background}

        \draw[gray,fill=gray,opacity=0.65](1.8,2.2) to[curve through={(0,4)}](1.8,5.8);
        \filldraw[fill=gray, opacity=0.65, draw=gray] (2.2,6.2) rectangle (5.8,9.8);
        \draw[gray,fill=gray,opacity=0.65](1.8,10.2) to[curve through={(0,12)}](1.8,13.8);
        
        \end{pgfonlayer}
        
        \begin{pgfonlayer}{background}
            \path [-,line width=0.1cm,black,opacity=1] (n0) edge node {} ($(n20)+(.05,0)$);
            \path [-,line width=0.1cm,black,opacity=1] (n1) edge node {} ($(n21)+(.05,0)$);
            \path [-,line width=0.1cm,black,opacity=1] (n2) edge node {} ($(n22)+(.05,0)$);
            \path [-,line width=0.1cm,black,opacity=1] (n3) edge node {} (n23);
            \path [-,line width=0.1cm,black,opacity=1] (n4) edge node {} (n24);

            \path [-,line width=0.1cm,black,opacity=1] (n0) edge node {} (n4);
            \path [-,line width=0.1cm,black,opacity=1] (n5) edge node {} (n9);
            \path [-,line width=0.1cm,black,opacity=1] ($(n10)+(-.05,0)$) edge node {} (n14);
            \path [-,line width=0.1cm,black,opacity=1] ($(n15)+(-.05,0)$) edge node {} (n19);
            \path [-,line width=0.1cm,black,opacity=1] ($(n20)+(-.05,0)$) edge node {} (n24);

            \node[fill=black,circle,draw=black,scale = 0.25] (b0) at (4,4) {};
            \node[fill=black,circle,draw=black,scale = 0.25] (b1) at (4,12) {};
            \node[fill=black,circle,draw=black,scale = 0.25] (b2) at (4,20) {};

            \node[fill=black,circle,draw=black,scale = 0.25] (b3) at (8,0) {};
            \node[fill=black,circle,draw=black,scale = 0.25] (b4) at (8,8) {};
            \node[fill=black,circle,draw=black,scale = 0.25] (b5) at (8,16) {};

            \node[fill=black,circle,draw=black,scale = 0.25] (b6) at (12,4) {};
            \node[fill=black,circle,draw=black,scale = 0.25] (b7) at (12,12) {};
            \node[fill=black,circle,draw=black,scale = 0.25] (b8) at (12,20) {};

            \node[fill=black,circle,draw=black,scale = 0.25] (b9) at (16,0) {};
            \node[fill=black,circle,draw=black,scale = 0.25] (b10) at (16,8) {};
            \node[fill=black,circle,draw=black,scale = 0.25] (b11) at (16,16) {};

            \node[fill=white,circle,draw=black,scale = 0.25] (a0) at (0,4) {};
            \node[fill=white,circle,draw=black,scale = 0.25] (a1) at (0,12) {};

            \node[fill=white,circle,draw=black,scale = 0.25] (a2) at (4,8) {};
            \node[fill=white,circle,draw=black,scale = 0.25] (a3) at (4,16) {};

            \node[fill=white,circle,draw=black,scale = 0.25] (a4) at (8,4) {};
            \node[fill=white,circle,draw=black,scale = 0.25] (a5) at (8,12) {};

            \node[fill=white,circle,draw=black,scale = 0.25] (a6) at (12,8) {};
            \node[fill=white,circle,draw=black,scale = 0.25] (a7) at (12,16) {};

            \node[fill=white,circle,draw=black,scale = 0.25] (a8) at (16,4) {};
            \node[fill=white,circle,draw=black,scale = 0.25] (a9) at (16,12) {};

            \node[fill=white,circle,draw=black,scale = 0.25] (a10) at (20,8) {};
            \node[fill=white,circle,draw=black,scale = 0.25] (a11) at (20,16) {};
    
        \end{pgfonlayer}
    \end{scope}

\end{tikzpicture}
}
\end{figure}
Starting with \begin{equation}
    \begin{split}
        \langle & X_{0},X_1, X_2, X_4, X_5, X_{2,3,6}, X_{2,3,7}, X_8, \\
        & Z_9, Z_{13}, Z_{14}, Z_{18}, Z_{19}, Z_{23}, Z_{24}\rangle \ ,
    \end{split}
\end{equation}
we arrive at
\begin{equation}
    \begin{split}
        \langle & (-1)^{a_0}Z_{0,1}, (-1)^{a_1}Z_{1,2,6,7}, (-1)^{a_2}Z_{2,3}, \\
        & \mathcolorbox{red!50}{X_{0,1,2,3}}, X_4, X_5, X_{2,3,6}, X_{2,3,7}, X_8, \\
        & Z_9, Z_{13}, Z_{14}, Z_{18}, Z_{19}, Z_{23}, Z_{24}\rangle.
    \end{split}
\end{equation}

\subsection{$X$ stabilisers}
We measure the yellow stabiliser in figure \ref{fig:injection_labelled_qb_reduced_X}.
\begin{figure}[!h]
\centering
\caption{\label{fig:injection_labelled_qb_reduced_X}}
\resizebox{0.5\linewidth}{!}{
    \begin{tikzpicture}[scale=.7,every node/.style={minimum size=1cm},on grid]
    \begin{scope}[yshift=0,every node/.append style={yslant=0,xslant=0},yslant=0,xslant=0,rotate=0]
        \fill[white,fill opacity=0.0] (0,0) rectangle (10,4);
        \node[fill=red!50,shape=circle,draw=black] (n4) at (2,18) {$4$};
        \node[fill=red!50,shape=circle,draw=black] (n3) at (2,14) {$3$};
        \node[fill=red!50,shape=circle,draw=black] (n2) at (2,10) {$2$};
        \node[fill=red!50,shape=circle,draw=black] (n1) at (2,6) {$1$};
        \node[fill=red!50,shape=circle,draw=black] (n0) at (2,2) {$0$};

        \node[fill=blue!50,shape=circle,draw=black] (n9) at (6,18) {$9$};
        \node[fill=red!50,shape=circle,draw=black] (n8) at (6,14) {$8$};
        \node[fill=red!50,shape=circle,draw=black] (n7) at (6,10) {$7$};
        \node[fill=red!50,shape=circle,draw=black] (n6) at (6,6) {$6$};
        \node[fill=red!50,shape=circle,draw=black] (n5) at (6,2) {$5$}; 

        \node[fill=blue!50,shape=circle,draw=black] (n14) at (10,18) {$14$};
        \node[fill=blue!50,shape=circle,draw=black] (n13) at (10,14) {$13$};

        \node[fill=blue!50,shape=circle,draw=black] (n19) at (14,18) {$19$};
        \node[fill=blue!50,shape=circle,draw=black] (n18) at (14,14) {$18$};

        \node[fill=blue!50,shape=circle,draw=black] (n24) at (18,18) {$24$};
        \node[fill=blue!50,shape=circle,draw=black] (n23) at (18,14) {$23$};

        \begin{pgfonlayer}{background}

        \filldraw[fill=yellow, opacity=0.65, draw=yellow] (6.2,14.2) rectangle (9.8,17.8);

        \draw[yellow,fill=yellow,opacity=0.65](10.2,18.2) to[curve through={(12,20)}](13.8,18.2);
        
        \filldraw[fill=yellow, opacity=0.65, draw=yellow] (14.2,14.2) rectangle (17.8,17.8);
        \end{pgfonlayer}
        
        \begin{pgfonlayer}{background}
            \path [-,line width=0.1cm,black,opacity=1] (n0) edge node {} ($(n20)+(.05,0)$);
            \path [-,line width=0.1cm,black,opacity=1] (n1) edge node {} ($(n21)+(.05,0)$);
            \path [-,line width=0.1cm,black,opacity=1] (n2) edge node {} ($(n22)+(.05,0)$);
            \path [-,line width=0.1cm,black,opacity=1] (n3) edge node {} (n23);
            \path [-,line width=0.1cm,black,opacity=1] (n4) edge node {} (n24);

            \path [-,line width=0.1cm,black,opacity=1] (n0) edge node {} (n4);
            \path [-,line width=0.1cm,black,opacity=1] (n5) edge node {} (n9);
            \path [-,line width=0.1cm,black,opacity=1] ($(n10)+(-.05,0)$) edge node {} (n14);
            \path [-,line width=0.1cm,black,opacity=1] ($(n15)+(-.05,0)$) edge node {} (n19);
            \path [-,line width=0.1cm,black,opacity=1] ($(n20)+(-.05,0)$) edge node {} (n24);

            \node[fill=black,circle,draw=black,scale = 0.25] (b0) at (4,4) {};
            \node[fill=black,circle,draw=black,scale = 0.25] (b1) at (4,12) {};
            \node[fill=black,circle,draw=black,scale = 0.25] (b2) at (4,20) {};

            \node[fill=black,circle,draw=black,scale = 0.25] (b3) at (8,0) {};
            \node[fill=black,circle,draw=black,scale = 0.25] (b4) at (8,8) {};
            \node[fill=black,circle,draw=black,scale = 0.25] (b5) at (8,16) {};

            \node[fill=black,circle,draw=black,scale = 0.25] (b6) at (12,4) {};
            \node[fill=black,circle,draw=black,scale = 0.25] (b7) at (12,12) {};
            \node[fill=black,circle,draw=black,scale = 0.25] (b8) at (12,20) {};

            \node[fill=black,circle,draw=black,scale = 0.25] (b9) at (16,0) {};
            \node[fill=black,circle,draw=black,scale = 0.25] (b10) at (16,8) {};
            \node[fill=black,circle,draw=black,scale = 0.25] (b11) at (16,16) {};

            \node[fill=white,circle,draw=black,scale = 0.25] (a0) at (0,4) {};
            \node[fill=white,circle,draw=black,scale = 0.25] (a1) at (0,12) {};

            \node[fill=white,circle,draw=black,scale = 0.25] (a2) at (4,8) {};
            \node[fill=white,circle,draw=black,scale = 0.25] (a3) at (4,16) {};

            \node[fill=white,circle,draw=black,scale = 0.25] (a4) at (8,4) {};
            \node[fill=white,circle,draw=black,scale = 0.25] (a5) at (8,12) {};

            \node[fill=white,circle,draw=black,scale = 0.25] (a6) at (12,8) {};
            \node[fill=white,circle,draw=black,scale = 0.25] (a7) at (12,16) {};

            \node[fill=white,circle,draw=black,scale = 0.25] (a8) at (16,4) {};
            \node[fill=white,circle,draw=black,scale = 0.25] (a9) at (16,12) {};

            \node[fill=white,circle,draw=black,scale = 0.25] (a10) at (20,8) {};
            \node[fill=white,circle,draw=black,scale = 0.25] (a11) at (20,16) {};
    
        \end{pgfonlayer}
    \end{scope}

\end{tikzpicture}
}
\end{figure}
Hence, we arrive at:
\begin{equation}
    \begin{split}
        \langle & (-1)^{a_0}Z_{0,1}, (-1)^{a_1}Z_{1,2,6,7}, (-1)^{a_2}Z_{2,3}, \\
        & \mathcolorbox{red!50}{X_{0,1,2,3}}, X_4, X_5, X_{2,3,6}, X_{2,3,7}, X_8, \\
        & (-1)^{b_0}X_{8,9,13,14}, (-1)^{b_1}X_{14,19},(-1)^{b_2}X_{18,19,23,24}, \\
        & Z_{9,13,19,24},\mathcolorbox{blue!50}{ Z_{9,14,19,24}}, Z_{18,24}, Z_{23,24}\rangle.
    \end{split}
\end{equation}

\subsection{Apply $T_4$}
We now apply $T_4$ gate to the initialised $\ket{+}_4$. Then measure the stabilisers: $X_4X_9$ and $Z_3Z_4Z_8Z_9$.
\begin{figure}[!h]
\centering
\caption{\label{fig:injection_labelled_qb_reduced_T_and_final_meas}}
\resizebox{0.5\linewidth}{!}{
    \begin{tikzpicture}[scale=.7,every node/.style={minimum size=1cm},on grid]
    \begin{scope}[yshift=0,every node/.append style={yslant=0,xslant=0},yslant=0,xslant=0,rotate=0]
        \fill[white,fill opacity=0.0] (0,0) rectangle (10,4);
        \node[fill=green!50,shape=circle,draw=black] (n4) at (2,18) {$T_4$};
        \node[fill=red!50,shape=circle,draw=black] (n3) at (2,14) {$3$};
        \node[fill=red!50,shape=circle,draw=black] (n2) at (2,10) {$2$};
        \node[fill=red!50,shape=circle,draw=black] (n1) at (2,6) {$1$};
        \node[fill=red!50,shape=circle,draw=black] (n0) at (2,2) {$0$};

        \node[fill=blue!50,shape=circle,draw=black] (n9) at (6,18) {$9$};
        \node[fill=red!50,shape=circle,draw=black] (n8) at (6,14) {$8$};
        \node[fill=red!50,shape=circle,draw=black] (n7) at (6,10) {$7$};
        \node[fill=red!50,shape=circle,draw=black] (n6) at (6,6) {$6$};
        \node[fill=red!50,shape=circle,draw=black] (n5) at (6,2) {$5$}; 

        \node[fill=blue!50,shape=circle,draw=black] (n14) at (10,18) {$14$};
        \node[fill=blue!50,shape=circle,draw=black] (n13) at (10,14) {$13$};

        \node[fill=blue!50,shape=circle,draw=black] (n19) at (14,18) {$19$};
        \node[fill=blue!50,shape=circle,draw=black] (n18) at (14,14) {$18$};

        \node[fill=blue!50,shape=circle,draw=black] (n24) at (18,18) {$24$};
        \node[fill=blue!50,shape=circle,draw=black] (n23) at (18,14) {$23$};

        \begin{pgfonlayer}{background}

        \filldraw[fill=gray, opacity=0.65, draw=gray] (2.2,14.2) rectangle (5.8,17.8);
        \draw[yellow,fill=yellow,opacity=0.65](2.2,18.2) to[curve through={(4,20)}](5.8,18.2);

        \end{pgfonlayer}
        
        \begin{pgfonlayer}{background}
            \path [-,line width=0.1cm,black,opacity=1] (n0) edge node {} ($(n20)+(.05,0)$);
            \path [-,line width=0.1cm,black,opacity=1] (n1) edge node {} ($(n21)+(.05,0)$);
            \path [-,line width=0.1cm,black,opacity=1] (n2) edge node {} ($(n22)+(.05,0)$);
            \path [-,line width=0.1cm,black,opacity=1] (n3) edge node {} (n23);
            \path [-,line width=0.1cm,black,opacity=1] (n4) edge node {} (n24);

            \path [-,line width=0.1cm,black,opacity=1] (n0) edge node {} (n4);
            \path [-,line width=0.1cm,black,opacity=1] (n5) edge node {} (n9);
            \path [-,line width=0.1cm,black,opacity=1] ($(n10)+(-.05,0)$) edge node {} (n14);
            \path [-,line width=0.1cm,black,opacity=1] ($(n15)+(-.05,0)$) edge node {} (n19);
            \path [-,line width=0.1cm,black,opacity=1] ($(n20)+(-.05,0)$) edge node {} (n24);

            \node[fill=black,circle,draw=black,scale = 0.25] (b0) at (4,4) {};
            \node[fill=black,circle,draw=black,scale = 0.25] (b1) at (4,12) {};
            \node[fill=black,circle,draw=black,scale = 0.25] (b2) at (4,20) {};

            \node[fill=black,circle,draw=black,scale = 0.25] (b3) at (8,0) {};
            \node[fill=black,circle,draw=black,scale = 0.25] (b4) at (8,8) {};
            \node[fill=black,circle,draw=black,scale = 0.25] (b5) at (8,16) {};

            \node[fill=black,circle,draw=black,scale = 0.25] (b6) at (12,4) {};
            \node[fill=black,circle,draw=black,scale = 0.25] (b7) at (12,12) {};
            \node[fill=black,circle,draw=black,scale = 0.25] (b8) at (12,20) {};

            \node[fill=black,circle,draw=black,scale = 0.25] (b9) at (16,0) {};
            \node[fill=black,circle,draw=black,scale = 0.25] (b10) at (16,8) {};
            \node[fill=black,circle,draw=black,scale = 0.25] (b11) at (16,16) {};

            \node[fill=white,circle,draw=black,scale = 0.25] (a0) at (0,4) {};
            \node[fill=white,circle,draw=black,scale = 0.25] (a1) at (0,12) {};

            \node[fill=white,circle,draw=black,scale = 0.25] (a2) at (4,8) {};
            \node[fill=white,circle,draw=black,scale = 0.25] (a3) at (4,16) {};

            \node[fill=white,circle,draw=black,scale = 0.25] (a4) at (8,4) {};
            \node[fill=white,circle,draw=black,scale = 0.25] (a5) at (8,12) {};

            \node[fill=white,circle,draw=black,scale = 0.25] (a6) at (12,8) {};
            \node[fill=white,circle,draw=black,scale = 0.25] (a7) at (12,16) {};

            \node[fill=white,circle,draw=black,scale = 0.25] (a8) at (16,4) {};
            \node[fill=white,circle,draw=black,scale = 0.25] (a9) at (16,12) {};

            \node[fill=white,circle,draw=black,scale = 0.25] (a10) at (20,8) {};
            \node[fill=white,circle,draw=black,scale = 0.25] (a11) at (20,16) {};
    
        \end{pgfonlayer}
    \end{scope}

\end{tikzpicture}
}
\end{figure}
After the application of $T_4$ we have:
\begin{equation}
    \begin{split}
        \rho' & \propto  \sum_{i,k} D_{i,k} (I+(-1)^{\delta_{k,Z}}X_4) \cdot \Lambda \ , \\
        D & = \begin{pmatrix}
       \text{cos}^2(\pi/8) I & (i/2)\text{sin}(\pi/4) Z_4 \\ 
       -(i/2)\text{sin}(\pi/4) Z_4  & \text{sin}^2(\pi/8) I
    \end{pmatrix} \ , \\
    \Lambda & = \prod_{j}(I+m_j) \\
    m_j \in   \langle & (-1)^{a_0}Z_{0,1}, (-1)^{a_1}Z_{1,2,6,7}, (-1)^{a_2}Z_{2,3}, \\
        & \mathcolorbox{red!50}{X_{0,1,2,3}}, X_5, X_{2,3,6}, X_{2,3,7}, X_8, \\
        & (-1)^{b_0}X_{8,9,13,14}, (-1)^{b_1}X_{14,19}, \\ 
        & (-1)^{b_2}X_{18,19,23,24}, \\
        & Z_{9,13,19,24},\mathcolorbox{blue!50}{ Z_{9,14,19,24}}, Z_{18,24}, Z_{23,24}\rangle \ .
    \end{split}
\end{equation}
If we perform the $Q = Z_3Z_4Z_8Z_9$ measurement, this implies:
\begin{equation}
    \begin{split}
        \rho' & \propto  \sum_{i,k} D_{i,k} (I+(-1)^{\delta_{k,Z}}\mathcolorbox{red!50}{X_{0,1,2,3,4}}) \cdot \Lambda \ , \\
        D & = \begin{pmatrix}
       \text{cos}^2(\pi/8) I & (i/2)\text{sin}(\pi/4) Z_4 \\ 
       -(i/2)\text{sin}(\pi/4) Z_4  & \text{sin}^2(\pi/8) I
    \end{pmatrix} \ , \\
    \Lambda & = \prod_{j}(I+m_j) \\
    m_j \in   \langle & (-1)^{a_0}Z_{0,1}, (-1)^{a_1}Z_{1,2,6,7}, (-1)^{a_2}Z_{2,3}, \\
    & (-1)^{a_3}Z_3Z_4Z_8Z_9, \\
        & X_5, X_{0,1,6}, X_{0,1,7}, X_{0,1,2,3,8}, \\
        & (-1)^{b_0}X_{8,9,13,14}, (-1)^{b_1}X_{14,19}, \\ 
        & (-1)^{b_2}X_{18,19,23,24}, \\
        & Z_{9,13,19,24},\mathcolorbox{blue!50}{ Z_{9,14,19,24}}, Z_{18,24}, Z_{23,24}, \rangle \ .
    \end{split}
\end{equation}
If we perform the measurement $Q = X_4X_9$, this implies (writing $\mathcolorbox{blue!50}{ Z_{9,14,19,24}} = \mathcolorbox{blue!50}{\bar{Z}}$ and $\mathcolorbox{red!50}{ X_{0,1,2,3,4}} = \mathcolorbox{red!50}{\bar{X}}$)
\begin{equation}
    \begin{split}
        \rho' & \propto  \sum_{i,k} D_{i,k} (I+(-1)^{\delta_{k,Z}}\mathcolorbox{red!50}{\bar{X}}) \cdot \Lambda \ , \\
        D & = \begin{pmatrix}
       \text{cos}^2(\pi/8) I & (i/2)\text{sin}(\pi/4) \mathcolorbox{blue!50}{ \bar{Z}} \\ 
       -(i/2)\text{sin}(\pi/4) \mathcolorbox{blue!50}{ \bar{Z}}  & \text{sin}^2(\pi/8) I
    \end{pmatrix} \ , \\
    \Lambda & = \prod_{j}(I+m_j) \\
    m_j \in   \langle & (-1)^{a_0}Z_{0,1}, (-1)^{a_1}Z_{1,2,6,7}, (-1)^{a_2}Z_{2,3}, \\
    & (-1)^{a_3}Z_3Z_4Z_8Z_9, \\
        & X_5, X_{0,1,6}, X_{0,1,7}, X_{0,1,2,3,8}, \\
        & (-1)^{b_0}X_{8,9,13,14}, (-1)^{b_1}X_{14,19}, \\ 
        & (-1)^{b_2}X_{18,19,23,24}, (-1)^{b_3}X_{4,9}\\
        & Z_{9,14,19,24}Z_{9,13,19,24}, Z_{18,24}, Z_{23,24}, \rangle \ .
    \end{split}
\end{equation}
Ignoring the other surface code stabilisers, which can be measured afterwards. We can see that vertical set of $\ket{+}$ initialised physical qubits ($0,1,2,3$) creates the $\bar{X}$ operator in the stabiliser decomposition superposition, when you measure the $Z$ type stabilisers. Similarly, the horizontally initialised qubits ($9,14,19,24$) induces the $\bar{Z}$ operator in the stabiliser decomposition superposition when you measure the $X$ type stabilisers. This shows the stabiliser decomposition approach can be used to analytically study magic state injection schemes with one $T$ gate. This can be generalised to arbitrarily high distance surface codes with a little bit of work.

\section{Post-selection}
Let's look at the stabiliser decomposition without the stabilisers (hide the other stabilisers in the object $\Lambda$):
\begin{equation}
    \begin{split}
        \rho' & \propto  \sum_{i,k} D_{i,k}(\mathcolorbox{blue!50}{Z_{4,9,14,19,24}}) (I+(-1)^{\delta_{k,Z}}\mathcolorbox{red!50}{X_{0,1,2,3,4}}) \Lambda \ .
    \end{split}
\end{equation}
Suppose you have experienced a Pauli error that flips an odd number of $X_{0,1,2,3,4}$. We have experienced a logical error. Similarly, if we have experienced a Pauli error that flips an odd number of $Z_{4,9,14,19,24}$, we have also experienced a logical error. In Li/Lao-Criger's schemes \cite{Li_2015,LaoInjection2022}, they post selections on all the stabiliser generators that is associated with these qubits. Hence, post selecting a state whereby none of these qubits are flipped an odd number of times to minimise the number of logical faults.

\section{Acknowledgments}
Kwok Ho Wan would like to thank Scott Aaronson and Daniel Gottesman for confirming a minor typo in their manuscript \cite{Aaronson_2004}\footnote{In the Physical Review A version of \cite{Aaronson_2004}, page 052328-12, in the third last equation on that page, the symplectic inner product in the exponent should contain $M_1$ instead of $M_j$.}. Kwok Ho Wan wants to thank his wife for telling him that 10:30 am minus 15 mins is equivalent to 15 mins past 10 am, this led to the realisation of $\pi/2-\pi/4 = \pi/4$. Which inspired the $R_Z(-\pi/4)\ket{Y} = \ket{T}$ representation in the $\alpha=1$ conditional $S$ gate proof for the $T$ gate by teleportation on $\ket{+}$. Kwok Ho Wan would also like to acknowledge discussions with Zhenghao Zhong (University of Oxford) on magic state injection in 2023 \cite{wan2025pauliwebsspuntransversal,wan2025pauliwebyranglestate}.

\bibliography{main}

\begin{thebibliography}{18}
\providecommand{\natexlab}[1]{#1}
\providecommand{\url}[1]{\texttt{#1}}
\expandafter\ifx\csname urlstyle\endcsname\relax
  \providecommand{\doi}[1]{doi: #1}\else
  \providecommand{\doi}{doi: \begingroup \urlstyle{rm}\Url}\fi

\bibitem[Gidney et~al.(2024)Gidney, Shutty, and Jones]{gidney2024magicstatecultivationgrowing}
Craig Gidney, Noah Shutty, and Cody Jones.
\newblock Magic state cultivation: growing {T} states as cheap as {CNOT} gates, 2024.
\newblock URL \url{https://arxiv.org/abs/2409.17595}.

\bibitem[Aaronson and Gottesman(2004)]{Aaronson_2004}
Scott Aaronson and Daniel Gottesman.
\newblock Improved simulation of stabilizer circuits.
\newblock \emph{Physical Review A}, 70\penalty0 (5), November 2004.
\newblock ISSN 1094-1622.
\newblock \doi{10.1103/physreva.70.052328}.
\newblock URL \url{http://dx.doi.org/10.1103/PhysRevA.70.052328}.

\bibitem[Lao and Criger(2022)]{LaoInjection2022}
Lingling Lao and Ben Criger.
\newblock Magic state injection on the rotated surface code.
\newblock In \emph{Proceedings of the 19th ACM International Conference on Computing Frontiers}, CF '22, page 113–120, New York, NY, USA, 2022. Association for Computing Machinery.
\newblock ISBN 9781450393386.
\newblock \doi{10.1145/3528416.3530237}.
\newblock URL \url{https://doi.org/10.1145/3528416.3530237}.

\bibitem[Fowler and Gidney(2019)]{fowler2019lowoverheadquantumcomputation}
Austin~G. Fowler and Craig Gidney.
\newblock Low overhead quantum computation using lattice surgery, 2019.
\newblock URL \url{https://arxiv.org/abs/1808.06709}.

\bibitem[Eastin and Knill(2009)]{eastin_knill_2009}
Bryan Eastin and Emanuel Knill.
\newblock Restrictions on transversal encoded quantum gate sets.
\newblock \emph{Phys. Rev. Lett.}, 102:\penalty0 110502, Mar 2009.
\newblock \doi{10.1103/PhysRevLett.102.110502}.
\newblock URL \url{https://link.aps.org/doi/10.1103/PhysRevLett.102.110502}.

\bibitem[Gottesman(2009)]{gottesman2009introductionquantumerrorcorrection}
Daniel Gottesman.
\newblock An introduction to quantum error correction and fault-tolerant quantum computation, 2009.
\newblock URL \url{https://arxiv.org/abs/0904.2557}.

\bibitem[Gidney(2021)]{gidney2021stim}
Craig Gidney.
\newblock Stim: a fast stabilizer circuit simulator.
\newblock \emph{{Quantum}}, 5:\penalty0 497, July 2021.
\newblock ISSN 2521-327X.
\newblock \doi{10.22331/q-2021-07-06-497}.
\newblock URL \url{https://doi.org/10.22331/q-2021-07-06-497}.

\bibitem[Gottesman(1998)]{gottesman1998heisenbergrepresentationquantumcomputers}
Daniel Gottesman.
\newblock The {H}eisenberg {R}epresentation of {Q}uantum {C}omputers, 1998.
\newblock URL \url{https://arxiv.org/abs/quant-ph/9807006}.

\bibitem[Bravyi and Gosset(2016)]{Bravyi_Gosset_2016_PRL}
Sergey Bravyi and David Gosset.
\newblock Improved {C}lassical {S}imulation of {Q}uantum {C}ircuits {D}ominated by {C}lifford {G}ates.
\newblock \emph{Phys. Rev. Lett.}, 116:\penalty0 250501, Jun 2016.
\newblock \doi{10.1103/PhysRevLett.116.250501}.
\newblock URL \url{https://link.aps.org/doi/10.1103/PhysRevLett.116.250501}.

\bibitem[Nakhl et~al.(2024)Nakhl, Harper, West, Dowling, Sevior, Quella, and Usman]{nakhl2024stabilizertensornetworksmagic}
Azar~C. Nakhl, Ben Harper, Maxwell West, Neil Dowling, Martin Sevior, Thomas Quella, and Muhammad Usman.
\newblock Stabilizer {T}ensor {N}etworks with {M}agic {S}tate {I}njection, 2024.
\newblock URL \url{https://arxiv.org/abs/2411.12482}.

\bibitem[Bremner et~al.(2010)Bremner, Jozsa, and Shepherd]{Bremner_2010}
Michael~J. Bremner, Richard Jozsa, and Dan~J. Shepherd.
\newblock Classical simulation of commuting quantum computations implies collapse of the polynomial hierarchy.
\newblock \emph{Proceedings of the Royal Society A: Mathematical, Physical and Engineering Sciences}, 467\penalty0 (2126):\penalty0 459–472, August 2010.
\newblock ISSN 1471-2946.
\newblock \doi{10.1098/rspa.2010.0301}.
\newblock URL \url{http://dx.doi.org/10.1098/rspa.2010.0301}.

\bibitem[Menczer et~al.(2024)Menczer, van Damme, Rask, Huntington, Hammond, Xantheas, Ganahl, and Legeza]{Menczer_NVIDIA_2024}
Andor Menczer, Maarten van Damme, Alan Rask, Lee Huntington, Jeff Hammond, Sotiris~S. Xantheas, Martin Ganahl, and {\"O}rs Legeza.
\newblock Parallel implementation of the density matrix renormalization group method achieving a quarter peta{FLOPS} performance on a single {DGX-H100} {GPU} node.
\newblock \emph{Journal of Chemical Theory and Computation}, 20\penalty0 (19):\penalty0 8397--8404, 2024.
\newblock \doi{10.1021/acs.jctc.4c00903}.
\newblock URL \url{https://doi.org/10.1021/acs.jctc.4c00903}.
\newblock PMID: 39297788.

\bibitem[Bayraktar et~al.(2023)Bayraktar, Charara, Clark, Cohen, Costa, Fang, Gao, Guan, Gunnels, Haidar, Hehn, Hohnerbach, Jones, Lubowe, Lyakh, Morino, Springer, Stanwyck, Terentyev, Varadhan, Wong, and Yamaguchi]{bayraktar2023cuquantumsdkhighperformancelibrary}
Harun Bayraktar, Ali Charara, David Clark, Saul Cohen, Timothy Costa, Yao-Lung~L. Fang, Yang Gao, Jack Guan, John Gunnels, Azzam Haidar, Andreas Hehn, Markus Hohnerbach, Matthew Jones, Tom Lubowe, Dmitry Lyakh, Shinya Morino, Paul Springer, Sam Stanwyck, Igor Terentyev, Satya Varadhan, Jonathan Wong, and Takuma Yamaguchi.
\newblock cu{Q}uantum sdk: A high-performance library for accelerating quantum science, 2023.
\newblock URL \url{https://arxiv.org/abs/2308.01999}.

\bibitem[Bravyi et~al.(2019)Bravyi, Browne, Calpin, Campbell, Gosset, and Howard]{Bravyi_2019}
Sergey Bravyi, Dan Browne, Padraic Calpin, Earl Campbell, David Gosset, and Mark Howard.
\newblock Simulation of quantum circuits by low-rank stabilizer decompositions.
\newblock \emph{Quantum}, 3:\penalty0 181, September 2019.
\newblock ISSN 2521-327X.
\newblock \doi{10.22331/q-2019-09-02-181}.
\newblock URL \url{http://dx.doi.org/10.22331/q-2019-09-02-181}.

\bibitem[Li(2015)]{Li_2015}
Ying Li.
\newblock A magic state’s fidelity can be superior to the operations that created it.
\newblock \emph{New Journal of Physics}, 17\penalty0 (2):\penalty0 023037, February 2015.
\newblock ISSN 1367-2630.
\newblock \doi{10.1088/1367-2630/17/2/023037}.
\newblock URL \url{http://dx.doi.org/10.1088/1367-2630/17/2/023037}.

\bibitem[Wan and Zhong(2025{\natexlab{a}})]{wan2025pauliwebyranglestate}
Kwok~Ho Wan and Zhenghao Zhong.
\newblock Pauli web of the {$|Y\rangle$} state surface code injection, 2025{\natexlab{a}}.
\newblock URL \url{https://arxiv.org/abs/2501.15566}.

\bibitem[ecz(2024)]{eczoo_bacon_shor_4}
\([[4,1,1,2]]\) four-qubit subsystem code.
\newblock In Victor~V. Albert and Philippe Faist, editors, \emph{The Error Correction Zoo}. 2024.
\newblock URL \url{https://errorcorrectionzoo.org/c/bacon_shor_4}.

\bibitem[Wan and Zhong(2025{\natexlab{b}})]{wan2025pauliwebsspuntransversal}
Kwok~Ho Wan and Zhenghao Zhong.
\newblock Pauli webs spun by transversal {$|Y\rangle$} state initialisation, 2025{\natexlab{b}}.
\newblock URL \url{https://arxiv.org/abs/2502.00957}.

\end{thebibliography}
\end{document}